\documentclass[preprint]{aastex}

\newfont{\curly}{cmsy10 scaled 1000}
\def\CR {\hbox{\curly{R}}}
\def\HI {\hbox{H {\sc i}}}
\def\HII {\hbox{H {\sc ii}}}
\def\etal   {{et~al.\/}}
\def\kms   {$km~s^{-1}$}
\def\moy   {$M_\odot~yr^{-1}$}
\def\Msun {M$_{\odot}$}

\begin{document}

\title{The Gas Content and Kinematics of Nearby Blue Compact Galaxies:  
Implications for Studies at Intermediate and High Redshift.}
\author{D.J. Pisano \& Henry A. Kobulnicky\altaffilmark{1}}
\affil{Astronomy Dept., U. Wisconsin - Madison}
\affil{475 N. Charter St., Madison, WI 53706}
\email{pisano@astro.wisc.edu, chip@astro.wisc.edu}
\author{Rafael Guzm\'an\altaffilmark{1}}
\affil{Dept. of Astronomy, University of Florida}
\affil{P.O. Box 112055, Gainesville, FL 32611}
\email{guzman@astro.ufl.edu}
\author{Jes\'us Gallego}
\affil{Departamento de Astrof\'{\i}sica, Universidad Complutense de Madrid}
\affil{E-28040 Madrid, Spain}
\email{jgm@astrax.fis.ucm.es}
\author{Matthew A. Bershady}
\affil{Astronomy Dept., U. Wisconsin - Madison}
\affil{475 N. Charter St., Madison, WI 53706}
\email{mab@astro.wisc.edu}
\altaffiltext{1}{Hubble Fellow}

\smallskip

\smallskip
\begin{abstract}

We present Arecibo 21 cm spectroscopy, Keck HIRES H$\beta$
spectroscopy, and WIYN R-band imaging of 11 nearby blue compact
galaxies with effective B-band surface brightness $SBe=$19.4-21.2 mag
arcsec$^{-2}$ and effective radii $R_{eff}=$0.6-1.9 kpc.  This sample
was selected to test the reliablility of mass estimates derived using
optical emission linewidths, particularly for the blue compact
star-forming galaxies observed at intermediate redshifts (0.1 $< z <$
1).  In addition, we also measure the \HI\ content and gas depletion
timescales for the nearby blue, compact galaxies in an attempt to
infer the present nature and possible future evolution of their
intermediate redshift analogs.  We detected \HI\ in 10 of 11 sample
galaxies.  They have \HI\ masses of 0.3--4$\times$10$^9~$M$_{\odot}$,
\HI\ linewidths, W$_{20}$, of 133--249 \kms, dynamical masses of
0.5--5$\times10^{10}$M$_{\odot}$, gas depletion timescales,
$\tau_{gas}$, of 0.3--7 Gyr, \HI\ mass fractions of 0.01--0.58, and
mass-to-light ratios of 0.1--0.8.  These values span the range of
values typical of nearby \HII\ galaxies, irregulars and spirals.
Despite the restricted morphological selection, our sample of galaxies
is quite heterogeneous in terms of \HI\ content, dynamical mass, and
gas depletion timescale.  Therefore, these galaxies have a variety of
evolutionary paths and should look very different from each other in 5
Gyr.  Those with high masses and gas depletion timescales are likely
to retain their ISM for future star formation while the lower mass
objects with small gas depletion timescales may be undergoing their
last major event of star-formation. Hence, the fading of
intermediate-redshift luminous blue compact galaxies into NGC205-type
spheroidals is a viable evolutionary scenario, but only for the least
massive, most gas-poor objects in this sample.

The most consistent characteristic of our morphologically-selected
sample is that the ratios of \HII\ linewidths to \HI\ 21 cm
linewidths, \CR=$\frac{W_{20}(\HII)}{W_{20}(\HI)}$, are systematically
less than unity, with an average value of \CR=0.66$\pm$0.16; similar
to findings for local \HII\ galaxies. The simplest explanation for
this result is that the ionized gas is more centrally concentrated
than the neutral gas within the gravitational potential.  We find that
\CR\ is a function of linewidth, such that smaller linewidth galaxies
have smaller values of \CR.  Correcting optical linewidths by this
factor not only raises the derived masses of these galaxies, but also
makes them consistent with the local luminosity--linewidth
(Tully-Fisher) relation as well.  If this ratio applies to
intermediate-redshift galaxies, then the masses of intermediate
redshift blue compact galaxies can be obtained from optical linewidths
after applying a small correction factor, and the proposed
luminosity evolution of the Tully-Fisher relation is much smaller and
more gradual than suggested by studies using optical emission line
width measurements.

\end{abstract}

\section{Introduction}

The Hubble Space Telescope and the new generation of 8 m class
telescopes have extended our knowledge of the early Universe by
identifying galaxies down to magnitudes B$\sim$28 and redshifts
z$\ge$3 (see review by Ellis 1997).  A current challenge is to
understand the evolutionary connection between distant galaxies and
their nearby counterparts.  One approach is to compare fundamental
galaxy parameters (i.e., sizes, masses, luminosities) of distant
samples with better-studied nearby counterparts to help understand the
harder-to-observe distant objects and establish evolutionary
connections to present-day galaxies.

Studies of high redshift galaxies have revealed a population of
compact, luminous galaxies with high star formation rates (e.g.,
Steidel, {\it{et al.}} 1996a; 1996b, Lowenthal {\it{et al.}} 1997).
At intermediate redshifts, apparently similar sources have been called
by various names, including ``Compact Narrow Emission Line
Galaxies'' (CNELG; Koo
\etal\ 1994; 1995; Guzm\'an \etal\ 1996, 1998), ``Blue Nucleated
Galaxies'' (BNG; Schade \etal\ 1995), and faint, compact galaxies
(Guzm\'an \etal\ 1997; Phillips \etal\ 1997) depending on what
properties are being emphasized, and how the samples were selected.
It is worth noting that the term ``Narrow'' in the context of CNELG is with
respect to QSOs and AGN; the CNELGs themselves display a range of
kinematic emission line-widths between 30 and 130 \kms.

Jangren \etal\ (2001) have compared several of the brighter of the
above intermediate-redshift samples. They find that while the sources
display a range of photometric properties, most of the galaxies can be
isolated quantitatively on the basis of their colors and image
structure parameters as a class distinct from normal Hubble-types
found in bright samples.  They define one such class as ``Luminous Blue
Compact Galaxies'' (LBCGs). Sources in this class have small sizes, high
luminosities (hence high surface-brightness), and very blue
colors. Their image concentration and asymmetry is slightly higher
than for nearby irregular galaxies included in bright samples.  
However, this class of galaxies is not defined by their emission-line
kinematic widths.

The nature and evolution of the LBCG class at intermediate redshift
are currently major issues under debate. The most comprehensive studies
of the LBCG population at intermediate redshift to date are those of
Phillips \etal\ (1997) and Guzm\'an \etal\ (1997). They concluded that
the LBCG class is populated by a mixture of starburst galaxies. About
60\% of galaxies in their sample are classified as ``HII-like''
since they are similar to today's population of luminous, young,
star-forming HII galaxies. The remaining $\sim$40\% are classified as
``SB disk-like'' since they form a more heterogeneous class of evolved
starbursts similar to local starburst spiral and irregular galaxies.
This classification is consistent with the results published for other
LBCG samples in the literature. For instance, Koo \etal\ (1994, 1995)
and Guzm\'an \etal\ (1996, 1998) first established the association
between HII galaxies and LBCGs for their sample of CNELGs.
Alternatively, Mall\'en-Ornelas \etal\ (1999) and Hammer \etal\
(2000), have concluded that their LBCG samples can be best identified
with bright irregulars, late-type spirals or even more massive spirals
with a very young bulge.

Given the diverse nature of the LBCG population, it is most likely
that they will not evolve into one homogeneous galaxy class.  Rather,
different LBCGs may evolve into different galaxy classes.
There are two main evolutionary scenarios
currently discussed in the literature. Koo \etal\ (1994) and
subsequent authors have suggested that some subset of the most compact
HII-type LBCGs at intermediate redshifts may be the progenitors of
local low-mass elliptical galaxies (also called spheroidals), such as
NGC 205. Their conjecture is based on the similarity of the kinematic
widths and sizes of the LBCGs with low-mass ellipticals. Evolutionary
models predict that, in 4-6 Gyr of passive evolution, the faded
luminosities and surface brightnesses of HII-type LBCGs will match
local low-mass ellipticals. This evolutionary prediction requires that
LBCGs are undergoing their last major burst of star formation at a
z$\sim$0.4. In order to match the low luminosities of low-mass
elliptical galaxies, star formation in HII-type LBCGs must be
short-lived (timescales $\sim$ 1 Gyr or less) so that they fade by
$\sim$2-4 magnitudes in $\sim$4-6 Gyr (Guzm\'an {\it{et al.}}  1998).
Since local low-mass ellipticals have little detectable cool gas
(Young \& Lo 1997; Young 2000), this prediction also requires HII-type
LBCGs to lose nearly all of their gas by the present
day. Alternatively, the evolutionary scenario for SB disk-like LBCGs
currently being considered is very different. Some authors have
suggested that these LBCGs may actually be disks forming from the
center outward to become present-day spirals (Phillips \etal\ 1997;
Hammer \etal\ 2000). Thus SB disk-like LBCGs would be more massive
than inferred from their virial masses since their small sizes
and kinematic emission-line widths would reflect mainly the central
starburst region (Phillips \etal\ 1997; Barton \& Van Zee 2001).
These objects are also expected to have large reservoirs of gas that
may allow for continuing star formation, albeit at a lower rate than
the current burst.

Measurements of the 21 cm neutral hydrogen (\HI) linewidth and flux
may provide an important new constraint on the evolution of LBCGs. The
\HI\ linewidth is a very good indicator of the rotational velocity of
a galaxy, provided the system is not severely distorted or
interacting.  We can use this rotational velocity to obtain a total
enclosed mass of the system, a parameter that will not change
dramatically with time (at least over the past few gigayears), unlike the 
star formation rate or luminosity.  \HI\ measurements typically trace gas 
out to a larger
radius than the optical emission lines and, hence, offer a more robust
measurement of the mass.  An integrated HI line flux, coupled with an
adopted distance, provides a measure of the total neutral hydrogen
mass which limits a galaxy's potential for future star formation. The
\HI\ mass and a known star formation rate (SFR) help set limits on the
timescale for the reservoir of neutral gas to be depleted.

Since neutral hydrogen observations of the intermediate redshift LBCGs
are not yet possible, we must infer their properties from nearby
analogs.  This is possible provided the nearby sample is
representative of the more distant LBCGs.  If observations imply that 
one of our analogs has a small
\HI\ linewidth, a small quantity of neutral gas, and a short gas
depletion timescale, then the case is strengthened
for it being a low-mass object
which will shortly cease star formation and undergo subsequent passive
evolution.  If, on the other hand, observations imply a large \HI\
linewidth, a large reservoir of neutral gas, and a long gas depletion
timescale, then the analog would appear to be a higher-mass galaxy.  A long
period of passive evolution following the end of the current starburst
would seem less likely and the galaxy may continue having subsequent
events of star formation.

Because we wish to consider sources which span a range of luminosities
from ``dwarf'' ($M_B>-18$) to ``luminous'' ($M_B<-18$), we adopt here
the more generic term Blue Compact Galaxy (BCG) to encompass this
diverse group of small, blue galaxies regardless of luminosity or
kinematics.  Throughout this paper, we will use the sample of compact
galaxies in the Hubble Deep Field flanking fields (HDF-ff; Guzm\'an
\etal\ 1997; Phillips \etal\ 1997), intermediate redshift CNELGs (Koo
\etal\ 1994, 1995; Guzm\'an \etal\ 1996), and the BNGs of Schade
\etal\ (1995) as a benchmark for assessing a population of local
compact galaxies.  We will refer to these previous samples
collectively as the intermediate-redshift BCGs.

Another issue relating to the study of intermediate redshift galaxies
that we will examine concerns the Tully-Fisher (T-F) relation and its
evolution. Specifically, studies of the internal kinematics of
intermediate redshift galaxies have led to discrepant results on the
evolution of the T-F zeropoint.  For example, Forbes \etal\ (1996),
Rix \etal\ (1997), Simard \& Pritchet (1998), and Mall\'{e}n-Ornelas
\etal\ (1999) all find between 1 and 2 magnitudes of brightening in
the T-F at redshifts of 0.2-0.8.  These surveys rely on [OII] or
[OIII] \& H$\beta$ emission-lines as kinematic tracers; all except
Simard \& Pritchet measure spatially integrated line-widths. In
contrast, studies using resolved rotation curves at intermediate
redshifts (Vogt \etal\ 1996, 1997; Bershady \etal\ 1999) find much
less brightening (0-0.6 mag) over a range of redshift between 0.1 to
1. Many of these studies specifically target blue galaxies, some of
which undoubtedly are BCGs by virtue of their strong emission lines
and high surface-brightness.

One question that has arisen is whether the spatially-integrated
line-widths of optical emission lines used in some studies
under-estimate the true rotation velocities.  Indeed, Forbes \etal\
(1996) note that in two cases where they have resolved rotation
curves, their line-widths under-estimate the true rotation speed by
2\% and 41\%, respectively. Complicating the issue is the fact that
the different studies select heterogeneously from the distant galaxy
population (Bershady 1997). If different galaxies evolve at different
rates, the measured evolution in the T-F zeropoint may depend on
galaxy types sampled, as suggested by Simard \& Pritchet (1998). They
find tentative evidence that the least-massive (slowest rotators)
brighten most. While they have measured rotation-curves (albeit at
limited spatial resolution and low S/N), the sense of this observation
is the same as would result from the systematic under-estimate of true
line-width based on optical measurements. We are in a position here to
address the reliability of internal kinematics based on
spatially-integrated, optical emission line-widths, as well as the
conclusions drawn from these measurements.

In this paper we are presenting \HI\ and \HII\ spectroscopy, as well
as R-band imaging, of 11 nearby BCGs drawn from the Universidad
Complutense de Madrid (UCM) emission-line survey (Zamorano {\it{et
al.}} 1994).  These galaxies were selected to morphologically resemble
the BCG population at intermediate redshift. In section 2 we present
the sample selection criteria, observing procedure, and reductions.
Section 3 contains the analysis of the data.  In section 4 we present
our sample galaxies.  We discuss the implications of our observations
for the inherent nature of BCGs in section 5, and we conclude in
section 6.  We adopt $H_0=70$ \kms\ $Mpc^{-1}$ and $q_0=0.05$
throughout this paper.

\section{Data \& Observations}

\subsection{Sample Selection}

Most BCGs at intermediate redshifts ($z\sim0.4$) that have been discussed 
in the literature are intrinsically
luminous objects ($M_B < -18$ mag), with blue rest-frame colors ($B-V
< 0.45, B-r < 0.6$), and compact in the sense that they have high
effective B-band surface brightnesses\footnote{$SB_e = 5\times
log[R_{eff} (kpc)] + M_B + 38.6$. This is the average surface brightness
inside the half-light radius, equivalent to the effective radius for a
r$^{1/4}$ profile.}  ($SB_e < 21.5$ mag arcsec$^{-2}$).  For this
study, we chose galaxies which would be nearby analogs to
intermediate-$z$ BCGs.  Sample galaxies were chosen from the UCM
emission line survey of star forming galaxies (Zamorano {\it{et al.}}
1994; 1996) based on data from the original survey, follow-up optical
spectroscopy (Gallego {\it{et al.}} 1996; 1997) and optical photometry
(Vitores {\it{et al.}}  1996a; 1996b; P\'erez-Gonz\'alez {\it{et al.}}
2000).  The survey preferentially selects objects with large H$\alpha$
emission line equivalent widths, $EW>50$\AA.  Four physical parameters
were used to identify the nearby counterparts to the objects observed
at intermediate redshift: luminosity, color, velocity dispersion
($\sigma$) of the optical emission lines, and effective radius. The
criteria were chosen to exclude very low luminosity systems and
Blue Compact Dwarfs, in the classical sense (M$_B\geq$ -17.0), to
include objects with blue color, $B-r~\leq$ 1.0, and those with small
sizes, R$_{eff}\leq$ 2.0 kpc, and small velocity dispersions,
$\sigma_{\HII}<$ 80 \kms. The important constraint is that our sample
galaxies have visible properties similar to BCGs at higher redshifts.
We further restricted the sample to objects with declinations between
-1\degr\ and +38\degr\ in order that they be observable from Arecibo.
From this limited sample we chose 11 galaxies, which tended to have
the smallest velocity dispersions.  Table~1 lists the targets selected
for observation, their positions, and the time-on-source, along with
alternate identifiers and comments.

In Figure~1 we compare the luminosity, color, size, and surface
brightness of the 11 UCM sample galaxies (filled circles) with
intermediate redshift BCGs from the Hubble Deep Field flanking fields
(open triangles; denoted HDF-ff; Phillips \etal\ 1997; Guzm\'an \etal\
1997) and with a sample of BCGs (open boxes, including the Compact
Narrow Emission Line Galaxies of Koo \etal\ 1994, 1995 and the Blue
Nucleated Galaxies of Schade \etal\ (1995)).  The polygons  (from
Phillips {\it{et al.}} 1997, figure 8) illustrate the approximate
location of nearby elliptical, spiral, irregular, and spheroidal
galaxies.   Figure~1 demonstrates that our sample galaxies have
similar surface brightnesses, absolute magnitudes, and colors to the
intermediate redshift BCGs from the literature.  In addition, our
galaxies have  surface brightnesses more similar to low-mass
ellipticals and high-mass  spheroidals than to massive spirals or
irregulars.  Some of the targets are on  the border between these two
classes of objects.  While our galaxies do not  occupy the full range
of parameter space that the intermediate redshift BCGs do,  they are
sampling a significant subset of the BCG properties.  Our galaxies
cover almost the full range of luminosities and colors as the
intermediate-z  BCGs, but are only comparable to the smallest of the
distant BCGs (with SB$_e\sim$20 mag arcsec$^{-2}$).  These limitations must be
remembered when we compare our sample galaxies with their more distant
counterparts.

\subsection{\HI\ Observations}

HI observations of the 11 sample galaxies were made using the 
Arecibo\footnote{The Arecibo Observatory is part of the National Astronomy 
and Ionosphere Center which is operated by Cornell University under a 
Cooperative Agreement with the National Science Foundation.}
305 m telescope on 1999 October 14-16.  We observed from 2200-0400 LST
each day using the 1.4 GHz feed in conjunction with the L-narrow and
L-wide receivers.  The L-wide receiver was only used for observations
of UCM 0135+2242, while  the L-narrow receiver was used for the
remaining 10 galaxies.  This configuration has a beamwidth of
3.$^{\prime}$4$\times$3.$^\prime$8 (79$\times$88 kpc for a typical
galaxy distance of 80 Mpc) and a measured gain of 10 K
Jy$^{-1}$.  The system temperature has been measured at 32 K for the
L-narrow receiver and 38 K for the L-wide receiver.  We observed in 4
IFs, recording data in two separate bandwidths in both  polarizations
simultaneously.  We observed with bandwidths of 12.5 MHz and 25  MHz
with 9-level sampling.  This yields a velocity coverage of $\sim$2500 
\kms and $\sim$5000 \kms, respectively.  With 2048
channels for  each IF, the velocity resolution is 1.3 \kms\ and
2.6 \kms, respectively.  Because the different subcorrelators
are not fully  independent, we could not combine the two separate
bandwidths to improve the sensitivity of our observations.

The observing procedure involved looking at the galaxy for 6 minutes
and then doing an off-source scan over the same zenith angle and
azimuth range for an additional 6 minutes.  We repeated observing
these on-off pairs until we felt we had a firm detection, or that no
detection was easily forthcoming.   Time on source and measured noise
levels for each target are presented in Tables 1 \& 2 respectively.

\subsection{\HI\ Reductions}

Calibration of the data was done using the Arecibo package, ANALYZ.
We started by combining each pair of on-off scans by taking
$\frac{on-off}{off}$.  Each combined pair was then corrected for gain
and  T$_{sys}$ variations with zenith angle and converted from T$_{B}$
in Kelvins  to flux in Janskys using the tabulated gain factors from
the Arecibo Users' Manual.  This factor is roughly 3.6 Jy K$^{-1}$,
but varies with the zenith angle of the  observation.  All scans for a
given galaxy were then averaged together.  We then  averaged the two
polarizations for each galaxy together to get a combined  un-polarized
spectrum.  Finally, we did a first or second order
baseline subtraction on the unpolarized,  combined spectrum to get
our final, calibrated, reduced spectrum of flux versus heliocentric
velocity.  As a check on our reductions, we compared the total \HI\ flux 
with previously  published values for UCM 2325+2318, otherwise known as 
NGC 7673.  We find  that the \HI\ flux agrees to within 5\% of the value
found by Nordgren {\it{et al.}} (1997a) with the VLA.  The data was then 
ported into IDL for the analysis.  Figure~2 shows the calibrated \HI\ 
spectra for all observed galaxies as the light grey lines.

\subsection {Keck Echelle Spectroscopy \& Reductions}

At the Keck I 10 m telescope\footnote{Some of the data presented
herein were obtained at the W.M. Keck Observatory, which is operated
as a scientific partnership among the California Institute of
Technology, the University of California and the National Aeronautics
and Space Administration.  The Observatory was made possible by the
generous financial support of the W.M. Keck Foundation.} we used the
HIRES spectrograph (Vogt {\it{et al.}} 1994) with the blue cross
disperser to obtain $R\sim$30,000 spectra over the wavelength range
3600 - 5200 \AA.  Observing occurred 1999 October 14-15 and 2000
September 14-15.  The primary goal was to use the H$\beta$
$\lambda$4861 emission line to measure the kinematics of the ionized
gas.  For two galaxies, UCM 2351+2321 and UCM 2325+2318, the redshifted
H$\beta$ line fell at the edge of the detector, so we instead use the
[O~III] $\lambda$4959 line in the following analysis.   The slit
decker D1 measuring 1.15\arcsec\ in the spectral direction and
14.0\arcsec\ in the spatial direction was used for all objects except
UCM0135+2242, UCM0148+2123, and UCM0159+2354 where the C5 slit decker
measuring 1.15\arcsec$\times$7.0\arcsec\ was used.  Periodic exposures
of a Th-Ar arc lamp were used to establish the radial velocity and
dispersion.  The wavelength zeropoint of each exposure is good to an
RMS of 0.003 \AA.  The mean instrumental resolution was 0.12 \AA\ FWHM
(7.4 \kms) at 4900 \AA\ based on exposures of arc lamps.  One
1200 s exposure was obtained for each object.  During each exposure,
the telescope was moved in order to drift the slit slowly across each
target to simulate a ``spatially integrated'' spectrum that would be
obtained if all objects were unresolved at large distances.  Inspection of the
resulting spectra showed that the 14\arcsec\ slit was long enough to
cover the entire emission line region of each galaxy, except in the
case of UCM2325+2318 where portions of the low-surface brightness disk
extended beyond the ends of the slit.  In the
reduction process, each spectrum was summed along the spatial
dimension to produce a 1D emission line spectrum.  The resulting
optical emission line spectrum for each galaxy is shown overplotted in
black on the \HI\ spectrum in Figure 2.  There is good correspondence between
the \HI\ and \HII\ systemic velocities, confirming that we have correctly
identified objects placed at the center of the Arecibo beam.  $W_{20}$
has been calculated from the \HII\ line for each galaxy observed
and the results appear in Table 2.  A comparison of our spatially integrated
spectra with the single position spectra of Gallego \etal\ (2001, in 
preparation) show that our values are larger, on average, by 7\% with an RMS
deviation of 19\%.  Therefore the linewidths of our galaxies are statistically
the same regardless of whether they are measured at a single position or are 
spatially-integrated.

\subsection {WIYN Imaging}

We obtained R-band images of all 11 galaxies with the
WIYN\footnote{The WIYN Observatory is a joint facility of the
University of Wisconsin-Madison, Indiana University, Yale University,
and the National Optical Astronomy Observatories.} 3.5 m telescope
using the Mini-Mosaic 4k$\times$4k CCD camera during 2000 October 16-17 and 
November 17-18 in 1.2\arcsec\ (mean) seeing.  The pixel scale was
0.141 arcsec/pixel, and integration times were 1-2 $\times300$ s.  The 
images were reduced in the standard manner.  We
use these images below to characterize the morphology of each galaxy
and supplement the data available in the literature (e.g., Vitores
\etal\ 1997; Gallego \etal\ 1996).  Rotational asymmetries, A, were
computed following the procedures and method specified in Conselice
\etal\ (2000); half-light radii, R$_{eff}$, and concentration indices, C, 
were computed following the procedures specified by Bershady \etal\
(2000). These values are therefore on the same system as computed for
the Frei (1999) local galaxy sample by these authors.

Because the half-light radii were in the range of 1.5 to 9 arcsec,
there were cases when we needed to apply corrections to the asymmetry
values, half-light sizes, and concentration indices to account for the
seeing. The corrections are in the sense of
increasing the asymmetry and image concentration, while decreasing the
half-light radius. The corrections for asymmetry are based on the
simulations presented in Figure 19 of Conselice \etal\ (2000), and
generally resulted in changes $<$ 15\%. However, we note that the
simulations included only one source plausibly similar to these
sources: NGC 4449.  To correct the half-light sizes and concentration
indices, we modeled the effects of seeing based on a simulated grid of 
aberrated, two-dimensional analytic light distributions (r$^{1/4}$-law
and exponential profiles with a range of axial ratios).  The corrections to 
the half-light radii were generally under 15\% and the changes in the 
concentration indices were typically a few tenths in the index.  
A comparison of our half-light radii and 
concentration indices with previous determinations show agreement of the 
means to 1\% and 28\% with an RMS deviation of 21\% and 12\%, respectively, 
to Vitores \etal\ (1996) and 6\% and 1\% for the mean with RMS deviation 
of 21\% 17\% to Pe\'rez-Gonz\'alez \etal\ (2000), with few outliers.  
Our adopted WIYN values for R$_{eff}$, C, and A are listed in Table 3.  For 
all subsequent discussions we will use these values, as determined from the 
WIYN images.  

\section{Analysis}

For each galaxy we measured the total \HI\ flux, the 20\% velocity width,
and the  central velocity of each galaxy as listed in Table 2.  The
total \HI\ flux was determined by integrating under the spectrum from where
the galaxy first appears from the noise, to where it becomes
indistinguishable from the noise.  The RMS was measured from the
off-line region and used to determine the error on the integrated
flux.  For UCM 2351+2321, which was not detected, the 3$\sigma$ upper
limit to the flux is set for an assumed \HI\ velocity width equal to
the optical width as noted in Table 2.  The optical velocity widths
come from our Keck observations.  As most of the galaxy spectra rise
very slowly out of the  noise, we Hanning smoothed the spectra to
assist in determining the 20\% velocity width.  Typically, we Hanning
smoothed over 3 or 5 channels, then we used a program to identify all
crossings of the 20\% of peak intensity.  We selected the most probable
velocity width from these options.  The cited velocity center is
simply the midpoint between the 20\% crossings.  The errors on these
values represent the approximate range of widths (and central
velocities) due to the uncertainty determining the 20\% crossing
points.  For UCM 2351+2321 the optical recession velocity is cited,
and used to calculate a distance, due to the lack of a detection in
\HI.  This distance is listed in Table 3 and is used to calculate
distance-dependent quantities such as \HI\ mass, luminosity, and
linear size.

We have also attempted to derive the dynamical masses for these
galaxies.   To do this we need a rotation velocity and a radius
corresponding to that  velocity.  We used half of the
inclination-corrected W$_{20}$ as a reasonable approximation for the 
rotation velocity.  For the corresponding radius, R$_{\HI}$, we
scaled the effective radius, R$_{eff}$, determined from our WIYN
images, up by a factor of 4.5.  This factor is based on the ratio of
the R$_{eff}$ to R$_{24.5}$ found for these galaxies in Vitores
{\it{et al.}} (1996), which  was 2.4.  We rounded this up to 2.5 to
get an estimate of R$_{25}$, and then multiplied R$_{25}$ by 1.8 to
get R$_{\HI}$, based on the observed ratio of R$_{25}$ to R$_{\HI}$
found by Broeils (1992) for nearby gas-rich galaxies.  R$_{\HI}$ may
even be higher as van Zee, Skillman, \& Salzer (1998) find
R$_{\HI}$/R$_{25}$ ranging from 2.8-4.9 for a sample of 5 \hbox{H {\sc
ii}} galaxies.  Using R$_{\HI}$ and V$_{rot}$, we calculate an
estimate for the dynamical mass:

\begin{equation}
M_{dyn}=\frac{V_{rot}^2 R_{\HI}}{G}.
\end{equation}

Table~4 lists these dynamical masses along with the hydrogen gas mass
fraction, f$_{gas}$=M$_{\HI}$/M$_{dyn}$.  Tables 3 and 4 also list the
optical properties of the sample galaxies, including absolute blue
magnitudes, M$_B$, B-r and B-V colors computed within the 24 mag
arcsec$^{-2}$ isophote, the effective surface brightness within the 24
mag arcsec$^{-2}$ isophote, the H$\alpha$ luminosity, and the
inclination of these galaxies (from Vitores \etal\ 1996a,
1996b;  Gallego \etal\ 1996, 1997; Pe\'rez-Gonz\'alez \etal\
2000).

To assist in examining the evolutionary potential of these galaxies we
calculated the star formation rates for these galaxies from the listed
redenning-corrected H$\alpha$ luminosities (Gallego {\it{et al}} 1996)
using the expression given in Kennicutt (1983):

\begin{equation}
 SFR (M_\odot~yr^{-1}) =\frac{L_{H\alpha}}{1.12\times 10^{41}~erg~s^{-1}} 
\end{equation} 

Using the star formation rate and the \HI\ mass of each galaxy, we calculate 
the gas depletion timescale, $\tau_{gas}=M_{HI}/SFR$ listed in Table 4.  
These estimates do not account for recycling of the gas, for the presence of 
helium ($\sim40$\% of the \HI\ mass), or for molecular gas.  All of these 
factors would increase $\tau_{gas}$\footnote{If we use the updated 
expression from Kennicutt {\it{et al.}} (1994) the SFR will be lower by 
$\sim$10\% ($\tau_{gas}$ will be higher by the same amount).  
The expression from Alonso-Herrero {\it{et al.}} (1996) will decrease the 
SFR and increase $\tau_{gas}$ by a factor of 2.8.}.  It is important to note 
that the L$_{H\alpha}$ values from Gallego \etal\ (1996) were measured 
through the spectrograph slit.  Therefore they only represent a fraction of 
the total L$_{H\alpha}$ of the galaxy as would be measured via narrow-band 
imaging.  Correcting for this underestimate would decrease our estimates of 
$\tau_{gas}$.  It is unclear, however, how to determine the extent of this 
correction with our current data.

Images of the target galaxies appear in Figure~3 and 4.  Figure 3
shows Digital Sky Survey (DSS) images centered on the observed target
coordinates.  Each image is 3.$^\prime$8 on a side, to match the
angular size of Arecibo beam and assist in determining the nature of the
objects we observed.  Figure~4 shows the WIYN R-band images of our
sample galaxies using a logarithmic greyscale.  In order to facilitate
comparison between galaxies, we have scaled these images so that each
dimension covers a projected distance of 10 kpc at the adopted
distances listed in Table~3.

\section{Galaxy Properties}

\subsection{UCM 0014+1829} 
Optically, UCM0014+1829 has an oblong central region with a  more
diffuse outer disk.  There is a bright star very close to the nucleus
which complicates the determination of its optical properties.  No
other galaxies are evident in the DSS image.  UCM 0014+1829 has the second
largest half-light radius, 1.7 kpc, in our sample, and is one of the
more concentrated objects as well, C=4.08.  It is relatively symmetric, 
A=0.08.  The
\HI\ profile is single peaked and slightly asymmetric.  The \HII\
spectrum is symmetric, centered at the same velocity as the peak of
the \HI\ profile, but is narrower with a width of 117 \kms\ vs
204 \kms.  A low amplitude wing to low velocity is evident in
both spectra.

\subsection{UCM 0040+0220}
UCM 0040+0220 is the featureless point source in the middle of the DSS
image.  It has a small size (R$_{eff}$=0.6 kpc), is symmetric 
(A=0.09), and is not very concentrated (C=3.0).  
This galaxy has one of the smallest $\tau_{gas}$
values, 0.41 Gyr, of our sample due, in part, to its low M$_{\HI}$,
2.6$\times$10$^8$\Msun.   While there is a strong double-horned \HI\
profile at  $\sim$4500 \kms, UCM 0040+0220 is the weaker
feature at $\sim$5400 \kms. The other features at higher
recession velocities may be real, but we do not have the
signal-to-noise ratio in this observation to confirm them. The \HI\
profile is somewhat asymmetric, but is hard to characterize due to its
low signal-to-noise ratio.  The \HII\ spectrum has a tail to lower
velocity,  similar to that in the \HI\ profile.  The double-horned
\HI\ profile at $\sim$4500 \kms\ is most likely the diffuse
galaxy east of UCM  0040+0220 in the DSS image.

\subsection{UCM 0056+0043}  
Our WIYN image of UCM 0056+0043 reveals that it is slightly extended
in the east-west direction.  No other galaxies are evident in the DSS
image of the field.  This galaxy has a half-light radius of  0.9 kpc,
an average concentration of 2.96, and a low asymmetry of 0.07.   The
\HI\ spectrum of UCM 0056+0043 is a strong single peak with weak high
velocity wings extending to lower redshift.  The profile is slightly
asymmetric.  The \HII\ spectrum matches the \HI\ profile very
closely, with an absence of high-velocity wings.  The \HII\ width
(126 \kms) is nearly identical to the \HI\ width (133 km
s$^{-1}$).  This galaxy has the smallest width of all our sample
galaxies, the smallest dynamical mass (5$\times$10$^9$\Msun), and the
highest gas mass fraction at 0.58.  It also has one of the longest gas
depletion timescales, 6.3 Gyr.

\subsection{UCM 0135+2242}  
This galaxy has a rather extended, diffuse disk evident in its WIYN
image with a large R$_{eff}$=1.4 kpc.  The concentration is high at 4.0, 
and the galaxy is very symmetric, A=0.02.  UCM 0135+2242 is one of the 
most massive galaxies in our sample, M$_{dyn}$=4.4$\times$10$^{10}$\Msun, 
and has a long $\tau_{gas}$ of 2.5 Gyr, despite its high SFR of 1.7 \moy.  
A potential small companion is evident just to the southwest in the image as
well.  The DSS image reveals another diffuse galaxy at the southern
most extent of the  Arecibo field, which may or may not be associated.
UCM 0135+2242 has a  double-horned \HI\ profile which is slightly
asymmetric.   The \HII\ profile is narrower (W$_{20}$=225 \kms)
and more  symmetric than the \HI\ profile (W$_{20}$=249 \kms).
Intriguingly, the low redshift peak of the \HI\  profile is broader
and stronger than the \HII\ peak.  In other words, the peak of the
\HI\  distribution does not contain  corresponding \HII\ emission.
This asymmetry in the \HI\ profile, and discrepancy with the \HII\
profile, may indicate that multiple galaxies were observed in the 
Arecibo beam, and/or that an interaction has occurred.

\subsection{UCM 0148+2123}
UCM 0148+2123 either has a double nucleus, or is two related objects which
are possibly merging based on our WIYN image of the galaxy.  No other 
galaxies are evident in the Arecibo beam.  The morphological
parameters suggest that this galaxy is not particularly large,
R$_{eff}$=1.0 kpc, nor concentrated, C=2.72, but it is quite asymmetric,
A=0.13.  The latter point is not surprising given the double-lobed
appearance of this galaxy.  In addition, this galaxy has one of the
smallest $\tau_{gas}$, 0.65 Gyr, meaning it will rapidly use up its
3.8$\times$10$^8$\Msun\ of \HI.  This could  be indicative of an
ongoing interaction.  UCM 0148+2123 has a asymmetric single-peaked
\HI\ profile, which may be caused by two galaxies being observed.  As
with most of our galaxies, the \HII\ profile is narrower (W$_{20}$=98
\kms) and more symmetric than the \HI\ profile (W$_{20}$=200 \kms).  
The \HI\ profile extends further to lower velocities than
the \HII\ profile does, suggesting a difference in the neutral and
ionized gas distributions

\subsection{UCM 0159+2354}
UCM 0159+2354 features an elongated core and extended, diffuse disk
indicative of a highly inclined system.  There may be some small,
faint galaxies also in this field, but it is not at all clear if they
are associated with our target galaxy.  While UCM 0159+2354 has a
normal R$_{eff}$, 1.0 kpc, it is the most concentrated galaxy in our
sample, 4.25, and is also optically asymmetric, A=0.12.  Also, UCM
0159+2354 is the reddest galaxy in our sample with a B-r color of
1.00.   The \HI\ spectrum shows a single-peaked profile for this
galaxy.  The center of the \HII\  spectrum is offset in velocity from
the \HI\ profile (4925 \kms\ vs. 4901 \kms).   The \HII\
profile is also narrower (138 \kms\ vs. 192 \kms).
This may indicate a real difference in the distributions of the
neutral and ionized gas.

\subsection{UCM 2251+2352}
There are two galaxies apparent in the \HI\ spectrum of UCM 2251+2352.
While both detections have similar widths, the one at higher redshift
is significantly brighter and is our desired target, based on its
coincidence with the \HII\ profile.  The \HI\ profile is wider than
the \HII\ profile  (140 \kms\ vs. 79 \kms), mostly due to
high-velocity wings,  while the central profiles match quite well.
The Digitized Sky Survey image shows a few faint galaxies to the south
and west of our target galaxy, but it is not clear if this is what we
have detected at lower redshift in our \HI\ spectrum.   Our WIYN
R-band image reveals a very faint, nearly face-on, barred spiral
structure.  The face-on nature  of this galaxy makes any dynamical
mass determination highly uncertain, due to the large inclination
corrections needed.  UCM 2251+2352 has a somewhat  extended disk, with
R$_{eff}$=1.4 kpc, and is relatively concentrated, C=3.43.  It is,
however, the most asymmetric galaxy in our sample, with A=0.27.

\subsection{UCM 2304+1640}
UCM 2304+1640 has a bright core, with a faint, extended, diffuse disk
surrounding it.  The DSS image of the field shows multiple diffuse
objects, suggesting that we may  have a spectrum of a galaxy group.
The half-light radius of UCM 2304+1640 is one  of the smallest in our
sample at 0.7 kpc.  The galaxy is rather concentrated, C=3.35, and
asymmetric, A=0.12, as well.  This galaxy has a single-peaked \HI\
profile which is somewhat rounded on top; this would be consistent
with observing a group of galaxies as seen in the DSS image.  
The \HII\ profile is more symmetric, narrower, and  slightly offset 
from the peak of the \HI\ profile, also consistent with multiple 
objects being in the Arecibo beam.

\subsection{UCM 2325+2318}
Both the WIYN and DSS images of UCM 2325+2318, also known as
NGC 7673, clearly shows
signatures of interactions, with loops and a tail apparent around the
main galaxy.  This is the most extended galaxy in our sample,
R$_{eff}$=1.9 kpc, and not very concentrated, C=2.98.  UCM 2325+2318 is
also the most asymmetric galaxy we observed, with A=0.60.  Conselice, 
Bershady \& Gallagher (2000) measured the same A value based on earlier
WIYN imaging, and noted this galaxy has a large ratio of W$_{20}$ to
W$_{50}$.  These authors found that a large value of this ratio was correlated
with morphology, indicative of recent interactions.  Consistent with this, 
Gallego {\it{et al.}} (1996) have identified three components in their \HII\ 
spectrum
of this galaxy and stated that this is an interacting system.  Nordgren \etal\ 
(1997a) observed the \HI\ disk of this system to be disturbed.  NGC 7673 is
also  a well-known luminous FIR source
(Sanders \& Mirabel 1996); not surprising considering that the SFR is
23.5 \moy, yielding $\tau_{gas}$=0.18 Gyr.  This is the highest SFR and
shortest $\tau_{gas}$ in our sample.  UCM 2325+2318 has a very strong,
single-peaked \HI\  profile. which has been studied in detail by
Homeier \& Gallagher (1999) with spectra from the WIYN DensePak
integral field unit.  In particular, they have  measured the velocity
field of this galaxy in the plane of the sky yielding a better
estimate of its rotation velocity, W$_{20}$=126 \kms\
vs. W$_{20}$=119 \kms\ from our  Keck spectrum; a fine
agreement.  The \HII\ profile is narrower than the \HI\ profile
(W$_{20}$=119 \kms\ vs. 164 \kms), and the relative
positions of the  peaks differ by $\sim10$ \kms.

\subsection{UCM 2329+2512}
The WIYN image of UCM 2329+2512 reveals a very small (R$_{eff}$=0.6
kpc), concentrated (C=4.15) galaxy with a faint, stellar loop,
possibly indicative of a recent interaction.  Despite this signature of
interaction, it is remarkably symmetric, A=0.02.  The low asymmetry 
is because the measurement of asymmetric is only in the central 
7$\arcsec$, and is dominated by the bright central region, and not the faint 
loops.  The faint loops could just be the projection of a strong two-armed 
spiral pattern in a highly inclined disk, and not a tidal tail.  The symmetric,
double-horned \HI\ profile would support the idea that this system has not 
been disturbed.  UCM 2329+2512 also has a low SFR, and the longest 
$\tau_{gas}$, 7.3 Gyr, of our sample.  No other galaxies are apparent 
in the DSS image of the Arecibo field.  The \HII\ spectrum of this galaxy is 
completely different from the \HI\ profile, 
with a symmetric single-peaked profile.  This suggests that the limited 
star formation in this galaxy is taking place only in the central 
regions, while the \HI\ is much more extended.

\subsection{UCM 2351+2321}
We did not detect \HI\ in UCM 2351+2321.  The 3$\sigma$ upper limit on
its  \HI\ mass is 1.3$\times$10$^8~$M$_{\odot}$ for an assumed width
160 \kms\ (from the \HII\ spectrum).  There is a hint of \HI\
emission  near where the \HII\ profile rises at high velocity, but it
is too weak to be a definite detection.  Optically, UCM 2351+2321 is
one of the reddest (B-r = 0.94  mag)  and most concentrated, C=4.02,
targets.  In some respects, this is the most interesting galaxy of the
entire sample because of the strong H$\alpha$ emission, yet weak \HI\
emission, implying a very small gas depletion timescale of $<$0.03
Gyr.  If any galaxy among our sample represents a transition type
between a gas-rich starburst and a gas-poor spheroidal,  this may be
such an object.

\section{Discussion}

\subsection{Kinematics of Local Blue Compact Galaxies}

The most interesting feature of our sample is that the ratio of \HII\ to 
\HI\ 21-cm linewidths, \CR=$W_{20}(\HII)$/$W_{20}(\HI)$, is always below
unity.  A further examination of the observed \HII\ and \HI\ profiles in 
Figure 2 shows that all but one of the \HII\ spectra (UCM 0135+2242) and all
but two of the \HI\ spectra (UCM~0135+2242 and UCM~2329+2512) are
single peaked, indicative of either extremely face-on disks or solid
body rotation curves commonly seen in \HII\ galaxies (e.g., van Zee
\etal\ 1998).  The latter option is more likely due to the large
velocity widths of the lines, despite being singly-peaked.

Figure 5 plots \CR\ versus the \HII\ emission
linewidths, $W_{20}(\HII)$.  Filled circles denote galaxies from this
paper, while crosses denote the spiral and irregular galaxies studied
by Kobulnicky \& Gebhardt (2000), triangles denote the spiral sample
of Raychaudhury \etal\ (1997), and squares are the \HII\ galaxies from
Telles \& Terlevich (1993).  It is evident from the figure that the
\HII--\HI\ linewidth ratio systematically departs from unity for small
\HII\ linewidths.  This effect is most prevalent for galaxies with
W$_{20}$(\HII)$\le$ 200 \kms.  The compact galaxies from our paper
show a systematic difference between the neutral and ionized gas
kinematics, such that they have a mean \CR=0.66$\pm$0.16, which is
different from most normal galaxies with a ratio closer to one.  Our
galaxies are similar, in this respect, to the \HII\ galaxies and \HII\
regions studied by Telles \& Terlevich (1993) which have linewidth
ratios $\sigma_{[O~III]}/\sigma_{\HI}\sim$0.7.  Figure 5 illustrates
that this discrepancy between the \HII\ and \HI\ linewidths seems to
vary smoothly from large linewidths to small. Therefore, we have
attempted to quantify this effect for all galaxies through an
approximate polynomial fit to the data as follows:

\begin{equation}
\CR=\frac{W_{\HII}}{W_{\HI}}=1-5~W_{\HII}^{-1}-500~W_{\HII}^{-2}-2.5\times10^5~W_{\HII}^{-3}
\end{equation}

This equation yields a very rapid drop-off in \CR\ below W$_{20}$(\HII) = 
100 \kms, which becomes unphysical below W$_{20}$(\HII) = 68 \kms.  Since 
we have not observed \CR\ values below 0.4, we suggest that when applying 
the above formula, a lower bound of 0.3 is imposed.

Note that there is still a large scatter around this fit, so this
formula should be applied with great caution.  A likely interpretation of the 
data is that the optical emission lines trace a smaller portion of the
gravitational potential than the \HI\ does, particularly in those
galaxies with the smallest linewidth galaxies (i.e., $\sigma <$40 \kms). 
For these galaxies, the dynamical masses based on the width
of optical emission lines will underestimate their masses by factors
of 2-4. However, for galaxies with velocity widths $\sigma$ larger
than $\sim$40 \kms\ the effect on the mass estimates is only
$<$20\%.

In addition to the observed dependence on velocity width, we have
attempted to find a set of observational parameters which can identify
galaxies in which the linewidth ratio, \CR, is small.  While we
searched for correlations between all the observed parameters (color,
SFR, concentration, asymmetry, R$_{eff}$, SB$_e$, etc.) and the
linewidth ratio, we had limited success.  In Figure 6 we plot the
relation between the linewidth ratio, H$\alpha$ equivalent width, and
concentration index; NGC 4449 is plotted as a comparison object.  Only
the lower panel shows any hint of a correlation: objects with higher
EW(H$\alpha$) seem to have lower linewidth ratios, while those lower
EW(H$\alpha$) have a wider range of ratios.  
The relation, really an upper envelope, between
EW(H$\alpha$) and \CR\ may indicate that galaxies undergoing a large
starburst have small linewidth ratios. Physically, this could
be caused by a minor merger driving some gas into the center of a
galaxy, triggering a central starburst (e.g. Hernquist \& Mihos 1995).
Therefore, most of the line emission will be centrally concentrated,
while the \HI\ disk will be more extended.  Such a starburst cannot
dominate the optical light, however, since there is no correlation
between the linewidth ratio or EW(H$\alpha$) and concentration index.
Alternatively, some of the \HI\ gas could be in the process of being
ejected from the galaxy and is no longer tracing the gravitational
potential.  This would require very large amounts of mass to be
ejected without being ionized.  Furthermore, the shapes of most of the
\HI\ profiles can be more simply explained by galactic rotation
than by ejection.  

\subsection{Neutral Gas Masses and the Potential For Future Star Formation}

The 10 galaxies with 21 cm detections have \HI\ masses ranging from
0.26-4.3$\times $10$^9~$M$_{\odot}$.  This \HI\ mass range is typical
for later-type spiral (Sc-Sm) and irregular galaxies (Roberts \&
Haynes 1994).  The SFRs of our 10 sample galaxies range between
0.23-23.5 M$_{\odot}$ yr$^{-1}$.  Thus, our samples spans a wide range
of \HI\ masses and H$\alpha$ luminosities (or SFR) similar to the
heterogeneous sample of spiral galaxies studied by Kennicutt, Tamblyn,
\& Congdon (1994; KTC) as shown in Figure 7.  The gas depletion
timescales for our sample galaxies are also similar to the KTC sample,
with typical $\tau_{gas}$ $>$ 1 Gyr, also shown in Figure 7.  Only
3-4 galaxies in our sample have $\tau_{gas}$ less than 1 Gyr: UCM
0040+0220, UCM 0148+2123, UCM 2325+2318, and UCM 2351+2321.  These are
not anomalously low for KTC spirals, but are on the lower edge of the
distribution.  The cause of these low $\tau_{gas}$ values is either
having a small \HI\ mass and a moderate SFR (e.g. UCM 0040+0220 and
UCM 0148+2123), or a high SFR and moderate M$_{\HI}$ (e.g. UCM
2325+2318), or a combination of high SFR and low M$_{\HI}$ (e.g. UCM
2351+2321).

From the inclination-corrected \HI\ linewidths, we estimate the
dynamical mass of each system using $R_{\HI}$ as a fiducial radius.
The masses range from $3.7\times10^{9}~M_\odot$ to
$1.5\times10^{10}~M_\odot$.  It should be noted that the derived
inclinations can be highly uncertain ($\pm$15$\degr$), and produce
large corrections to the \HI\ linewidths for low inclinations.  Such
errors in inclination propagate into our mass determination as the
square of the linewidth.  If our estimate of $R_{\HI}=4.5~R_{eff}$ is
too large, then the dynamical masses should be reduced in proportion to
$R_{\HI}$.

Using our derived M$_{dyn}$ and calculated M$_{\HI}$ we can calculate
the neutral hydrogen gas mass fraction, f$_{gas}$.  The gas mass
fractions for our sample have reasonable values that range from 0.01
to 0.54 with a mean value of 0.12.  Thus, some of these galaxies may
be gas rich like nearby \HII\ and irregular galaxies
($f_{gas}\sim0.3-0.5$; Van Zee, Salzer \& Skillman 1998) while others
could have gas mass fractions typical of the Milky Way and other
spirals ($f_{gas}<0.10$; Roberts \& Haynes 1994).

There are some ambiguities in the interpretation of \HI\ profiles
worth consideration.  Our estimates of \HI\ masses may be too large if
there are multiple galaxies in the relatively large Arecibo beam (90
kpc in extent at an average distance of 83 Mpc).  Multiple galaxies
are not necessarily easy to identify from a spatially integrated \HI\
profile.  They can produce asymmetric single or doubly peaked
profiles, and/or flat or rounded-top \HI\ profiles (see Nordgren
{\it{et al.}} 1997a, 1997b for examples).  Depending on the masses and
dynamics of the system, the asymmetries may be quite small, even if
the relative masses are high.  An overestimate of the \HI\ masses
would result in an overestimate of $\tau_{gas}$.  Therefore, these
galaxies may be using up their \HI\ at a more prodigious rate than
indicated by the present estimates.  Finally, it is difficult to
determine if interactions are ongoing in the sample galaxies from the
\HI\ profiles.  Interactions will affect the observed linewidths, and, hence, 
the accuracy of our calculated dynamical masses.  
Strong single peaks are predicted and seen in some
merger remnants (Horellou \& Booth 1997; Bendo \& Barnes 2000), but
other interacting systems or merger remnants are indistinguishable
from normal, isolated galaxies (Horellou \& Booth 1997).  Only with
higher spatial resolution \HI\ observations can these ambiguities be
cleared up.  For this paper we can only assume that we have observed
single galaxies.

\subsection{Comparison to Intermediate Redshift Blue Compact Galaxies}

Over the last few years, there has been an increasing controversy
surrounding the reliability of the virial mass estimates using
velocity widths measured from ionized gas emission in BCGs at
intermediate redshifts. It is well known that the ionized gas does not
always sample the whole range of the rotation curve, since the space
distribution of the ionized gas is typically more compact than the
extended HI (e.g., Broeils 1992; Taylor et al. 1995).  Velocity widths 
derived using ionized gas emission lines likely underestimate the
actual galaxy gravitational potential, and thus the inferred virial
mass. The key question is:  ``How do we relate optical line-widths 
to the rotation velocity?"  Since emission line velocity
widths are often used to estimate galaxy masses in unresolved BCGs
at higher redshifts, it is important to assess the degree at which
these measurements may be biased by a likely underestimation of the
actual rotation velocity.

For BCGs at redshift $z<0.35$, the issue was first discussed by
Guzm\'an \etal\ (1996) who re-analyzed \CR\ in the published data for a
sample of nearby HII galaxies (Telles \& Terlevich 1992) and concluded
that ``since CNELGs are similar to H II galaxies, the measured
velocity widths are likely to underestimate their internal velocities
by as much as 30\%''.  A 30\% increase was thus applied to velocity
widths measurements in subsequent papers (e.g., Guzm\'an \etal\ 1998).
The value of this correction was also investigated by Rix \etal\
(1997) who compared the velocity dispersion in the ionized gas,
$\sigma_v$, with the maximum circular velocity of the ionized gas,
$V_c$, for three, well-resolved, nearby galaxies.  They found an average
value of $\sigma_v/V_c=0.6$ which is consistent with the analytic
prescription of Tully \& Fouqu\'e (1985). On the other hand, Lehnert
\& Heckman (1996) showed a very poor correlation between emission line
velocity widths and rotational velocities for a sample of local
starburst galaxies (see their figure 13). They noted that
emission lines from nuclear regions trace only a fraction of the
rising portion of the rotation curve in nuclear starbursts, but they
did not investigate how this affected the spatially-integrated optical
spectrum of the galaxy, or how global optical linewidths compared to
the \HI\ linewidths. Neither Rix \etal\ (1997) nor Lehnert \& Heckman
(1996) discussed the relation of the ionized gas kinematics to the neutral
gas kinematics, which is the issue of interest here. Kobulnicky \&
Gebhardt (2000) performed integrated optical spectroscopy of 21
galaxies, including some of the Lehnert \& Heckman (1996) nuclear starburst
sample, and they found good agreement between optical and \HI\
linewidths for all objects except NGC~4449. Note that NGC~4449 is the
object in the Kobulnicky \& Gebhardt (2000) sample that most closely 
resembles the BCG population.

In order to assess the relevance of our work for distant galaxies, it
is important to consider how analogous our sample of 11 galaxies is to
the intermediate redshift BCGs that we wish to learn more about.  We 
can examine how our sample galaxies compare in the
M$_B$--$\sigma_{\HII}$ and $R_{eff}$--$\sigma_{\HII}$ planes.  In
Figure 8 we compare the distribution of our sample galaxies to the
BCGs of Phillips \etal\ (1997) and Guzm\'an \etal\ (1998).  While our
galaxies span nearly the full range of M$_B$ and R$_{eff}$ as the
distant BCGs, they have systematically smaller linewidths. Indeed, the
nearby sample shows a mean emission line velocity width near 30 \kms\
compared to 50-60 \kms\ for the BCGs; both sets of linewidths are
uncorrected for inclination.  Figure 8 brings into question whether we
are truly observing analogs to the intermediate redshift BCGs. We
believe, however, that our sample galaxies, while not representative
of all intermediate redshift BCGs, are similar to those with the
smallest linewidths (i.e., $\sigma<$ 40 \kms).

The hypothesis that we are observing analogs to at least some of the
intermediate redshift BCGs is supported by examining other
morphological and photometric properties of our galaxies. In Figure
9, we plot the $B-V$ colors versus asymmetry, $A$, concentration
index, $C$, and B-band surface brightness with the half-light radius,
$SBe$, for our sample galaxies, the Frei (1999) sample of local,
luminous elliptical and spiral galaxies with parameters measured by
Bershady \etal\ (2000), the luminous BCG sample from Jangren \etal\
(2001), and the BCG sample from the HDF-ff with parameters measured by
Guzm\'an \etal\ (1997).  With the exception of two red sources
(UCM0159+2354 and UCM2351+2321), our galaxies occupy a location in the
$B-V$ vs. $SBe$ and $B-V$ vs. $C$ planes that overlaps strongly with
the intermediate redshift BCG samples and is clearly distinct from
normal, nearby galaxies.  Most of the galaxies in our sample are less
luminous and redder than the majority of the intermediate redshift
BCGs.  Based on the definition of BCGs from Jangren \etal\ (2001)
7 of our 11 galaxies are ``BCGs'' (the exceptions are UCM 0148+2124,
UCM 0159+2354, UCM 2329+2512, and UCM 2351+2321).  Of those seven,
only the three most luminous galaxies in our sample technically fit
the ``LBCG'' definition of Jangren \etal\ (2001) (UCM 2325+2318, UCM
0135+2242, and UCM 2251+2352).  All of these points serve to illustrate
that we must be very careful about what we infer regarding the nature
of intermediate redshift BCGs based on observations of our sample of
galaxies or similar samples in the local universe.  For instance, the two 
nearby galaxies Barton \& van Zee (2001) propose are local counterparts to the 
intermediate redshift BCG population would not have been selected as such by
the color - surface brightness criteria defined by Jangren \etal\ (2001).  
These two nearby galaxies are either redder or of lower surface brightness 
(or both) than the intermediate redshift BCGs as well as the nearby sample 
studied here.  With this caveat in mind, we will now examine what we can 
learn about intermediate redshift BCGs.

At the top of Figure 5, we have drawn several boxes to represent the
range of linewidths for the galaxies in the studies of the Hubble Deep
Field flanking fields (Phillips \etal\ 1997), z$\sim$0.6 field
galaxies (Mall\'en-Ornelas \etal\ 1999), z$\sim$0.25 field galaxies
(Rix \etal\ 1997) and compact blue galaxies at z$\sim$0.4 (Guzm\'an
\etal\ 1998).  Figure 5 shows that the discrepancy between \HII\ and
\HI\ linewidths described in Section 4.1 could affect the mass
determinations in all these studies. The effect can be as much as a
factor of 2-4 for the smallest velocity width galaxies.  However,
according to our empirical relation, the effect may be only
$\sim$30-40\% for galaxies with velocity widths $\sigma >$ 40 km
s$^{-1}$.  Note that this effect is in addition to other
well-recognized uncertainties in mass determinations due to unknown
inclinations and clumpy gas distributions (modeled for spiral galaxies
by Rix \etal\ 1997). A more rigorous approach to apply the results
from our study to the intermediate redshift BCG population is to
restrict our conclusions only to the seven galaxies that can be
considered ``bona fide'' counterparts according to Jangren \etal's
definition. These galaxies have a \CR\ ranging from 0.54 to 0.94,
with a mean value of 0.70.  The three ``bona fide" LBCG counterparts
have essentially the same mean and range.  For the 3 non-BCG galaxies 
we detected in \HI\, the mean value of \CR\ is only 0.54.  These values of 
\CR\ for BCGs are in good agreement with the corrections applied to the 
mass estimates in previous work at higher redshift (e.g., Guzm\'an \etal\ 
1996, 1998).

	One of the most interesting conclusions derived using mass
estimates of BCGs at intermediate redshifts is that their
mass-to-light ratios are roughly one order of magnitude lower than
those values characteristic of local galaxies with similar
luminosities (Guzm\'an \etal\ 1996; Phillips \etal\ 1997). This
conclusion can be tested with our sample of nearby BCGs. For
consistency with previous work, we define the mass-to-light ratio
within R$_{eff}$ (instead of R$_{HI}$), using the velocity dispersion of 
the \HI\, and uncorrected by inclination effects.  
This definition allows a direct comparison with values
published in the literature for a wide variety of galaxy types. For
our entire sample, the derived mass-to-light ratios range from 0.07 to 0.78
M$_\odot$/L$^B_\odot$.  The mean mass-to-light ratio for non-BCGs, 0.6 
M$_\odot$/L$^B_\odot$, is higher than for the BCGs, 0.3 
M$_\odot$/L$^B_\odot$, in our sample.  These values are consistent with the
mass-to-light ratios for intermediate redshift BCGs studied by Phillips 
\etal\ (1997).  The mean absolute magnitude is 
M$_B = -19.2$ mag;  about 1.3 magnitudes fainter than L$\star$.  Using the same
representative sample of local galaxies in the RC3 (de Vaucouleurs
\etal\ 1991) analyzed by Phillips \etal\ (1997), we estimate that nearby, 
``normal'' galaxies with M$_B \sim -19.2$ have mass-to-light
ratios that approximately range from 0.8 to 10 M$_\odot$/L$^B_\odot$.  
The observed range in mass-to-light ratio for BCGs
inconsistent with that characteristic of massive systems with comparable
luminosities, such as Sa-Sc spiral galaxies. These spirals tend to be
the most abundant galaxy type with this luminosity in the local
universe. Instead, BCGs have mass-to-light ratios that are similar to
the lowest observed values in lower-mass galaxies of similar
luminosity, such as irregulars and HII galaxies.  As our BCG counterparts
evolve, their mass-to-light ratios will increase to values more comparable
to normal, spiral galaxies, provided that they fade.  

	Finally, the observed distribution of gas depletion timescales
$\tau_{gas}$ provides additional information about the nature and
evolution of the distant BCG galaxy population. Of the seven BCGs in
our sample, two have $\tau_{gas}<0.5$ Gyrs, two have $\tau_{gas}\sim1$
Gyrs, and three have $\tau_{gas}>2$ Gyrs. The wide range observed
suggests a very heterogeneous population of objects with very
different burst strengths and reservoirs of neutral gas. The BCGs with
the smallest gas depletion timescales will shortly cease star
formation and thus may be good candidates to undergo subsequent
passive evolution. However, a long period of passive evolution
following the end of the current starburst in BCGs with $\tau_{gas}>2$ 
Gyr seems less likely and these galaxies may continue having subsequent
events of star formation. 

\subsection{Implications for the Evolution of the Tully-Fisher Relation}

A related controversy exists concerning whether velocity widths of
ionized gas are suitable for the study of galaxy scaling relations
such as the Tully \& Fisher relation (1977; T-F).  Traditionally, the
T-F relation is a correlation measured between luminosity and HI
line-width. It has been well established that ``terminal'' rotation
speeds derived from spatially resolved velocity fields (in the optical
or HI) provide a well-defined surrogate for HI line-widths for normal
spiral galaxies in the local universe (e.g., Mathewson \etal 1992,
Courteau 1997). Ionized gas line-widths are of particular interest for
intermediate- and high-redshift work where spatial resolution and
signal-to-noise are limited and HI is out of reach with present
telescopes and instrumentation. However, there has been only limited
exploration of whether optical {\it line-widths} are suitable
surrogates for use in measuring the T-F relation (e.g., Rix \etal\
1997; KG00). Despite this fact, there have been a number of studies
attempting to exploit the observability of optical line-widths for
drawing inferences about the evolution of M/L of intermediate-redshift
galaxies, as discussed in the Introduction. Our purpose here is to
demonstrate that these results should be viewed with caution.

Figure 10 shows how our sample of compact galaxies compares to the T-F
relation.  Each galaxy in our sample is plotted three times. Dotted
circles denote the measured \HII\ linewidths, $W_{20}(\HII)$, circles
denote the measured \HI\ 21-cm linewidths, $W_{20}(HI)$, and filled
circles denote the inclination-corrected \HI\ 21-cm linewidths,
$W_{20}(HI)/sin(i)$.  For comparison, we also plot \HII\ galaxies from
Telles \& Terlevich (1993; crosses), and the B-band Tully-Fisher
relation of Pierce \& Tully (1988; solid
line).\footnote{$M_B=-6.86\log(V_M)-2.27$. In order to convert $V_M$
to the observed quantity, $W_{20}$, we adopt the prescription of Tully
\& Fouqu\'e (1985),
\begin{equation} 
(2\times V_M)^2 =W_R^2 = W_{20}^2 + W_t^2 -
2W_{20}W_t[1-e^{-(W_{20}/W_c)^2}] - 2W_t^2 e^{-(W_{20}/W_c)^2}.
\end{equation}
$W_R$ is the rotation full amplitude which is 2$\times V_{M}$.
$W_t=38$ \kms\ is the width due to turbulent motions and $W_c=120$
\kms\ is the transition point between galaxies having Gaussian and
those having double-horned \HI\ profiles.} While the compact galaxies
in our sample lie near the relation for \HII\ galaxies and show a
large offset from the canonical T-F relation when using observed \HII\
line-widths, if we we adopt the inclination-corrected \HI\ widths our
sample is consistent on average with the T-F relation for normal
galaxies.

Prompted by this result, we have re-examined the results of T-F
studies at intermediate redshift employing line-widths.  Specifically,
we have re-computed magnitude offsets from T-F for the Forbes \etal\
(1996), Rix \etal\ (1997), and Mall\'{e}n-Ornelas \etal\ (1999)
samples after applying the line-width correction in Equation~3 (Figure
5). The line-widths of the Forbes \etal\ sample are such that
corrections based on our formulation are minimal, despite the
discussion noted above based on their own rotation curve data. Indeed,
it is sobering to note that of the two galaxies for which they have
rotation curves, the galaxy with the small line-width correction
appears to have a bimodal light distribution, while the galaxy with
the large line-width correction appears to be a relatively normal
spiral with large luminosity and size (see their figure 1).

For Rix \etal\ we use their tabulated line-widths and adopt the Pierce
\& Tully (1988) T-F relation, which has a slope of $\gamma=-6.86$. The
adopted slope is within 1$\sigma$ of the value derived by Rix \etal\
(1997) for their intermediate-redshift sample; ours is a conservative
choice since their mean slope is much steeper ($\gamma=-14.3$) and
hence results in much larger magnitude-offset corrections. We find a
mean offset of -1.8 mag without applying line-width correction. This
is -0.3 mag larger than their quoted mean value, consistent with our
ignoring the color-dependence to the T-F zeropoint (see their \S3.3.1,
eqn. 9).  However, with line-width corrections we find a mean offset
of -1.1 mag, or +0.7 mag less brightening than estimated by Rix \etal\
(1997). Accounting for their color-dependence results in a net,
corrected brightening of -0.8 mag in the T-F zeropoint at $z\sim0.25$.

To make a fair re-assessment with Mall\'{e}n-Ornelas \etal 's results, we
adopt their T-F slope of $\gamma=-7.46$. Since Mall\'{e}n-Ornelas \etal\
(1999) do not provide tabulated data nor a zeropoint for their
fiducial T-F relation, we estimate a correction for their median
line-width (60 \kms) estimated from their figure 2, and calculate a
differential correction, $\Delta M_c$, given as:
\begin{equation}
\Delta M_c = -\gamma \log (\frac{W_{\HII}}{W_{\HI}}),
\end{equation}
where the linewidth ratio is given by our Equation~3.  We find a
median correction is +0.2 mag, although we note that at the lower
line-width limit of their sample the correction is +1.9 mag, which is
comparable (but opposite in sign) to their average estimated
brightening.

In isolation, the picture that emerges from this re-analysis of
intermediate redshift line-width studies is that the evolution of the
T-F zeropoint is more gradual than previously suggested -- even for
the bluest galaxies studied by Rix \etal\ (1997) and
Mall\'{e}n-Ornelas \etal\ (1999). For these galaxies we estimate -0.8
mag at $z\sim0.25$ and -1.8 mag at $z\sim0.6$. However even this
revised estimate should still be tempered by the fact that corrections
for color-dependence in the T-F relation relation and inclination are
uncertain. For example, the color-dependence of the T-F zeropoint
estimated by Rix \etal\ is $\sim$3 times smaller than what is reported
in Bershady \etal\ (2001). In the absence of direct inclination
measurements (e.g., b/a), Rix \etal\ (1997) have performed an elegant
estimate of the mean inclination of their sample.  They take into
account that the bluest galaxies may preferentially lie at relatively
low inclinations (where internal reddening is minimized). However, our
sample's mean inclination is 43$^\circ$, with a maximum of 60$^\circ$;
almost all of our sample is more face-on than the mean inclination of
$\sim57^\circ$ adopted by Rix \etal\ (1997). If our sample is
representative of the inclination distribution of the blue galaxy
population, taking into account this change in mean projection results
in +0.6 mag less brightening (assuming $\gamma=-6.86$). Similarly,
Mall\'{e}n-Ornelas \etal\ (1999) compare their uncorrected velocities
for intermediate redshift galaxies to a local sample from the
RC3. While they make identical cuts in b/a ($<0.8$), the two samples
may not have comparable inclination distributions. Our arguments here
are admittedly tentative, and are not intended to diminish the
pioneering efforts to measure the T-F relation at intermediate
redshifts.  Yet the striking fact remains that most
intermediate-redshift studies employing spatially-resolved rotation
curves find less evolution (e.g., Vogt \etal 1996, 1997; Bershady
\etal\ 1999), and that by using the inclination-corrected \HI\
linewidths even the galaxies studied in this paper fall squarely on
the local T-F relation.

A corollary to the result in Figure 10 is that our local sample of
BCGs cannot fade nor brighten dramatically {\it and} remain on the T-F
relation. The relatively large reservoir of gas in most of these
systems indicates these galaxies will likely experience star-formation
in their future, as discussed in previous sections. However the
amplitude of past and future star-formation is not well constrained,
and so the basic assumption that these systems remain (or have been)
on the T-F relation is weak. Analogous galaxies at intermediate
redshift also will be so weakly bound. Nonetheless, if we assume these
galaxies have remained on the T-F relation (or brighter) over
lifetimes in excess of 5 Gyr, then the current burst makes only a
small contribution to the total stellar mass. Consequently, in this
scenario these galaxies also are not candidates for strong fading in
the future.

However, it is important to stress that the extant characteristics of
the local galaxies studied here are dissimilar to the extreme LBCGs
for which a strong fading scenario was originally proposed by Koo
\etal\ (1995) and Guzm\'an \etal\ (1996). The fading these authors
postulated was part of a transformation process whereby some subset of
the LBCGs evolved into today's spheroidals. As such, these LBCGs would
not be expected to lie or remain on the T-F relation at any time. In
other words, the assumption that these galaxies must remain on the T-F
relation, made in the previous paragraph, is not valid in this
scenario.

Finally, the trends of our line-width ratio, \CR, with line-width also
has implications for the line-widths of Lyman Break galaxies (LBGs)
observed at z$>$2.7 (e.g., Pettini \etal\ 2001).  Pettini \etal\ are
skeptical of whether their ionized gas line-widths are tracing
mass. They note, for instance, that they find little evidence for a
line-width luminosity relation in their sample or in a larger sample
at z$\sim$1. The effect of the line-width dependence of \CR\ will be
to decrease the slope of the T-F relation, as shown in Figure 10.
While the effect is rather mild in Figure 10, it may be more extreme
for galaxies at higher redshift. It is striking that while the LBGs
are 1-3 magnitudes brighter than the intermediate redshift LBCGS, they
have comparable range of ionized-gas velocity widths, with mean values
of 70 and 60 \kms\ respectively. Indeed, the gas content of galaxies is
presumably larger at higher redshifts, and hence it may be reasonable
to expect that the value and trend of \CR\ with line-width evolves and
becomes more extreme at higher redshift.  We share, then, Pettini
\etal's caution in interpreting their ionized gas line-widths, but
remark that our corrections should at least provide a lower limit to
what should be applied to the ionized gas line-widths at high
redshift.

\section{Conclusions}

The present sample of compact galaxies were chosen to be analogous to
the Blue Compact Galaxies (BCGs) at intermediate redshift, so that we
could infer the gaseous properties of such galaxies.  For our sample,
we find that the neutral hydrogen masses, hydrogen gas mass fractions,
star formation rates, and gas depletion timescales are comparable to
nearby gas-rich spiral, irregular and HII galaxies.  The broad range
of HI properties imply BCGs form a very heterogeneous galaxy
class. Our sample galaxies have substantial amounts of neutral
hydrogen ($\sim$10$^9~$M$_{\odot}$) and large inferred total masses
($\sim$10$^{10}~$M$_{\odot}$).  However, unlike most nearby, luminous
spiral and irregular galaxies, the ratio of \HII\ to \HI\ linewidths, \CR\, is
systematically less than unity with a mean value near 0.66.  In this
respect our sample is similar to nearby \HII\ galaxies.  The magnitude
of \CR\ is close to what was assumed by Guzm\'an \etal\ (1998) and others 
for intermediate redshift BCGs.  In addition, 
\CR\ varies with \HII\ linewidth, departing dramatically from
unity for W$_{20}(\HII)\le$ 200 \kms.  We have parameterized this
relationship to help those who wish to attempt a correction in higher
redshift data.  Unfortunately, there is no tight correlation between
any of the morphological or photometric properties of these galaxies
and the linewidth ratio, \CR.  The upper limit of \CR\ does appear to
decrease with increasing H$\alpha$ equivalent width, suggesting that
this effect is related to central starbursts in these galaxies, but
this is tenuous conclusion.  Applying the corrections due to
inclination and converting from \HII\ to \HI\ linewidth for our
galaxies moves them onto the Tully-Fisher relation, leaving little room
for these galaxies to fade if they are to remain on the relation. 
This implies that the evolution of the Tully-Fisher relation from
intermediate redshift to the present day is smaller and more gradual
than has been previously been suggested (e.g., Rix \etal\ 1997).

Although the luminosities, colors, and effective sizes of our sample
were chosen to mimic those of the BCG population at intermediate
redshifts, our sample remains heterogeneous and only samples a portion of the
properties of BCGs.  Of the 11 galaxies we observed, only 7 can be classified 
as BCGs based on the definition of Jangren \etal\ (2001).  Only the brightest
4 of those 7 BCGs are classified as LBCGs.  Furthermore, our galaxies have 
systematically smaller linewidths than the observed population of intermediate
redshift BCGs. They
also tend to have, on average at a given luminosity, smaller sizes.
Therefore our sample is not representative of all intermediate
redshift BCGs, but only a subset of them.  As such, we are limited in
what we can infer regarding the properties of the distant BCGs. For
those intermediate redshift BCGs with small velocity widths and sizes,
we can assume that they will form also a heterogeneous population of
relatively gas rich galaxies ($\sim$10$^9~$M$_{\odot}$) whose gas
fractions ($f_{gas}=0.02-0.5$) and gas depletion timescales
($\tau_{gas}=0.2-6$ Gyrs) may span well over one order of magnitude.
Although it is not possible to constrain the evolutionary descendants
of these BCGs from the present data, we can remark that the relatively
large dynamical and \HI\ masses measured for about two-thirds of our sample
will function to inhibit the ejection of the ISM and to fuel future
star formation. Although the current bursts of star formation are
strong, they are unlikely to be the last, nor may they produce the
dominant stellar population in many of these galaxies. The remaining
one-third of our sample shows, however, gas depletion timescales short
enough to suggest that they may cease star formation in the near
future and thus may be good candidates to undergo subsequent passive
evolution.

Future work on this topic is essential.  Higher spatial resolution
\HI\ data will help us resolve any confusion regarding the
interpretation of our single-dish observations.  It will also allow us
to probe the internal structure of these galaxies to better understand
how the effects of interactions may be affecting our calculated
masses. Observations of the molecular gas in these galaxies will help
determine the total gas mass available for star formation in these
galaxies.  Finally, observations of a much larger sample which is more
representative of the BCG class at intermediate redshift is necessary
to better infer the properties of their distant counterparts.  This is
the only way to divine the gaseous properties of the intermediate
redshift BCGs until we are capable of observing them directly.

\section{Acknowledgments}

We wish to thank the staff at the Arecibo Observatory for all their
assistance and hospitality during our observations there.  They helped
make the observing run both enjoyable and successful.  Special thanks
to Snezana Stanimirovic, our ``friend'' at Arecibo, who went above and
beyond the call of duty in helping us prepare for observing, observe,
and reduce the data.  We thank P.G.  P\'erez-Gonzalez for providing B
magnitudes and B-r colors for our sample galaxies.  We also thank John
Salzer, Liese van Zee, and David Koo for their comments which helped improve 
the manuscript. The Digitized Sky Surveys were produced at the Space
Telescope Science Institute under U.S. Government grant NAG
W-2166. The images of these surveys are based on photographic data
obtained using the Oschin Schmidt Telescope on Palomar Mountain and
the UK Schmidt Telescope.  The Second Palomar Observatory Sky Survey
(POSS-II) was made by the California Institute of Technology with
funds from the National Science Foundation, the National Geographic
Society, the Sloan Foundation, the Samuel Oschin Foundation, and the
Eastman Kodak Corporation.  D.J.P.  acknowledges partial support from
the Wisconsin Space Grant Consortium.  H.A.K. is grateful for support
from Hubble Fellowship, HF-01094.01-97A, award by the Space Telescope
Science Institute.  R.G. acknowledges support from Hubble Fellowship,
HF-01092.01-97A, award by the Space Telescope Science Institute.
J.G. acknowledges partial support form the Spanish Programa Sectorial
de Promoci\'on del Conocimiento under grant PB96-0645.
M.A.B. acknowledges for support from NSF grant AST-9970780 and LTSA 
contract NAG5-6032.

\newpage

\begin{deluxetable}{lcccr}  
\tablecolumns{5}  
\tablewidth{0pc}  
\tablecaption{Targets}  
\tablehead{  
\colhead{Name} &\colhead{RA (1950)}   & \colhead{Dec. (1950)}    & \colhead{Obs. Time} & \colhead {Notes}   \\
\colhead{}     &\colhead{$^h:^m:^s$} & \colhead{$^o:^{\prime}:^{\prime\prime}$}   & \colhead{min.} & \colhead{}}
\startdata  
UCM0014+1829 & 00:14:39.9 & 18:29:38 & 18 & \nodata \\
UCM0040+0220 & 00:40:15.6 & 02:20:24 & 24 & UM 63 \\
UCM0056+0043 & 00:56:30.2 & 00:43:53 & 12 & UM 296 \\
UCM0135+2242 & 01:35:13.6 & 22:42:04 & 42 & \nodata \\
UCM0148+2124 & 01:48:18.5 & 21:23:53 & 54 & \nodata \\
UCM0159+2354 & 01:59:00.5 & 23:54:44 & 54 & \nodata \\
UCM2251+2352 & 22:51:18.9 & 23:52:14 & 18 & \nodata \\
UCM2304+1640 & 23:04:26.2 & 16:40:02 & 36 & \nodata \\
UCM2325+2318 & 23:25:11.9 & 23:18:49 & 24 & NGC 7673, merger? \\
UCM2329+2512 & 23:29:36.1 & 25:12:09 & 12 & \nodata \\
UCM2351+2321 & 23:51:06.8 & 23:21:16 & 48 & \nodata \\

\enddata  
\end{deluxetable}  

\begin{deluxetable}{lcrrrrr}  
\tablecolumns{7}  
\tablewidth{0pc}  
\tablecaption{Line Measurements}  
\tablehead{  
\colhead{Name} & 
\colhead{RMS}  & 
\colhead{Vel$_\odot$} & 
\colhead{W$_{20}(HI)$}   &
\colhead{$\int{S}dv$} &
\colhead{$W_{20}(\HII)$}  &
\colhead{\CR\tablenotemark{1}}  \\
\colhead{ } &
\colhead{mJy/chan} &   
\colhead{km~s$^{-1}$}   &   
\colhead{km~s$^{-1}$}   &   
\colhead{Jy~km~s$^{-1}$} & 
\colhead{km~s$^{-1}$} &
\colhead{ } 
}
\startdata  
UCM0014+1829 & 1.28 &  5249$\pm$3 & 204$\pm$5  & 1.15$\pm$0.05 &  117$\pm5$ & 0.57$\pm$0.02 \\
UCM0040+0220 & 1.04 &  5432$\pm$3 & 136$\pm$5  & 0.18$\pm$0.02 & 96$\pm5$  & 0.65$\pm$0.04 \\
UCM0056+0043 & 1.52 &  5352$\pm$3 & 133$\pm$5  & 2.1$\pm$0.1 &  126$\pm9$  & 0.94$\pm$0.07 \\
UCM0135+2242 & 1.11 & 10354$\pm$3 & 249$\pm$5  & 0.83$\pm$0.04 & 225$\pm15$ & 0.90$\pm$0.06 \\
UCM0148+2124 & 0.64 &  4890$\pm$10& 200$\pm$20 & 0.32$\pm$0.02 &  98$\pm8$ & 0.49$\pm$0.06 \\
UCM0159+2354 & 0.48 &  4901$\pm$3 & 192$\pm$5  & 0.52$\pm$0.01 &  138$\pm12$ & 0.71$\pm$0.06  \\
UCM2251+2352 & 1.12 &  7715$\pm$5 & 140$\pm$10 & 0.60$\pm$0.03 &  79$\pm9$ & 0.56$\pm$0.07 \\
UCM2304+1640 & 0.79 &  5162$\pm$3 & 157$\pm$5  & 0.99$\pm$0.02 &  85$\pm$6   & 0.54$\pm$0.04 \\
UCM2325+2318 & 1.07 &  3412$\pm$3 & 164$\pm$5  & 7.21$\pm$0.06 & 119$\pm7$\tablenotemark{a} & 0.72$\pm$0.04 \\
UCM2329+2512 & 1.24 &  3718$\pm$3 & 203$\pm$5  & 2.50$\pm$0.07 & 88$\pm6$ & 0.43$\pm$0.03 \\
UCM2351+2321 & 0.68 &  7943$\pm$3\tablenotemark{b}   & \nodata & $<0.042$\tablenotemark{c} & 165$\pm$15 &  \nodata \\
\enddata

\tablenotetext{1}{\CR= ${{W_{20}(\HII)}/{W_{20}(HI)}} $}
\tablenotetext{a}{An independent measurement by N. Homeier (private communication)
using WIYN DensePak H$\alpha$.
observations with a 30\arcsec$\times$45\arcsec\ integral field yields 126 $km~s^{-1}$.}
\tablenotetext{b}{Optical velocity from emission lines.}
\tablenotetext{c}{3$\sigma$ upper limit over 63 channels (165 \kms).}
\end{deluxetable}  
 
\begin{deluxetable}{lrrrrrrrrrr}  
\tablecolumns{11}  
\tablewidth{0pc}  
\tablecaption{Morphological Parameters}
\tablehead{  
\colhead{Name}&\colhead{Dist}&\colhead{M$_{B}$}&\colhead{B-r }&\colhead{B-V }&\colhead{$R_{eff}$}&\colhead{inc} &\colhead{SBe }&\colhead{EW(H$\alpha$)} &\colhead{C}  &\colhead{A}     \\  
\colhead{ }   &\colhead{($Mpc$)}&\colhead{($mag$)}&\colhead{($mag$)}&\colhead{($mag$)}&\colhead{($kpc$)} &\colhead{(deg)} &\colhead{($mag/sq$ \arcsec)}&\colhead{(\AA)}&\colhead{}&\colhead{}  \\
\colhead{ }   &\colhead{(1)}    &\colhead{(2)}             &\colhead{(3) }   &\colhead{(4)}    &\colhead{(5)}    &\colhead{(6)}     &\colhead{(7)} &\colhead{(8)} &\colhead{(9)} &\colhead{(10)} }       
\startdata  
UCM0014+1829 &  75 & -18.5 & 0.05 & 0.16  & 1.7 & 50 & 21.2 & 146 & 4.08 & 0.08 \\
UCM0040+0220 &  78 & -17.7 & 0.67 & 0.50  & 0.6 & 30 & 19.9 &  97 & 3.00 & 0.09 \\
UCM0056+0043 &  76 & -18.1 & 0.42 & 0.37  & 0.9 & 60 & 20.3 &  61 & 2.96 & 0.07 \\
UCM0135+2242 & 148 & -19.6 & 0.59 & 0.46  & 1.4 & 45 & 19.7 &  60 & 4.00 & 0.02 \\
UCM0148+2124 &  70 & -17.7 & 0.47 & 0.48  & 1.0 & 40 & 20.9 & 150 & 2.72 & 0.13 \\
UCM0159+2354 &  70 & -17.5 & 1.00 & 0.83  & 1.0 & 60 & 21.1 &  74 & 4.25 & 0.12 \\
UCM2251+2352 & 110 & -19.1 & 0.59 & 0.47  & 1.4 & 15 & 20.2 &  84 & 3.43 & 0.27 \\
UCM2304+1640 &  74 & -17.1 & 0.27 & 0.28  & 0.7 & 45 & 20.6 & 155 & 3.35 & 0.12 \\
UCM2325+2318 &  49 & -20.5 & 0.30 & 0.30  & 1.9 & 45 & 19.4 & 101 & 2.98 & 0.60 \\
UCM2329+2512 &  53 & -17.4 & 0.64 & 0.49  & 0.6 & 45 & 19.9 & 160 & 4.15 & 0.02 \\
UCM2351+2321 & 113 & -17.8 & 0.94 & 0.77  & 0.7 & 35 & 20.1 & 117 & 4.02 & 0.03 \\
\enddata  
\tablenotetext{1}{Using D(Mpc)=Velocity/$H_0$; $H_0=70~km~s^{-1}~Mpc^{-1}$. }
\tablenotetext{2}{Assumes $H_0=70$, $q_0=0.05$, from Vitores \etal\ (1996).}
\tablenotetext{3}{Observed B-r color from Vitores {\it{et al.}} (1996); typical uncertainties are 0.12 mag}
\tablenotetext{4}{B-V color derived from the observed B-r color, without correction
for redenning, assuming an
Irregular galaxy spectral energy distribution, and B-V=0.56(B-r)+0.136 based on Fukugita \etal\ (1995).}
\tablenotetext{5}{Effective radius in kpc from our WIYN photometry, corrected
for the effects of seeing, and assuming the above cosmology.}
\tablenotetext{6}{Inclination, derived from the eccentricity, $e=1-b/a$,
as in Aaronson, Huchra, \& Mould (1980), $i~(deg)=cos^{-1}
sqrt{(1.042*E^2-0.042)} + 3$ where E=1-e.  Typical uncertainties are
15$^\circ$.}
\tablenotetext{7}{Effective surface brightness, computed from $SB_e=5\times\log(R_{eff}(kpc))+M_B+38.6 $.}

\tablenotetext{8}{$H\alpha$ equivalent width in \AA\ (Gallego \etal\ 1996)}
\tablenotetext{9}{Concentration index from our WIYN photometry.}
\tablenotetext{10}{Asymmetry parameter from our WIYN photometry.}
\tablenotetext{a}{3$\sigma$ upper limit for 63 channels (165 \kms).}
\end{deluxetable}  
 
\begin{deluxetable}{lrrrrrrrrr}  
\tabletypesize{\small}
\tablecolumns{10}  
\tablewidth{0pc}  
\tablecaption{Derived Parameters}
\tablehead{  
\colhead{Name}&\colhead{M$_{\HI}$} &\colhead{L$_{H\alpha}$} & \colhead{SFR} & \colhead{ $\tau_{gas}$} &\colhead{$V_{rot}$} & \colhead{R$_{\HI}$} & \colhead{M$_{dyn}$} & \colhead{$f_{gas}$} & \colhead {M/L}   \\  
\colhead{ }   &\colhead{10$^9~$M$_{\odot}$}&\colhead{10$^{41}$ ergs s$^{-1}$}&\colhead{M$_{\odot}$/yr}&\colhead{Gyr}&\colhead{\kms} & \colhead{kpc} &\colhead{10$^{10}~$M$_{\odot}$} & \colhead{ } & \colhead{M$_\odot$/L$_{\odot}$} \\
\colhead{ }   &\colhead{(1)}  &\colhead{(2)} &\colhead{(3)} &\colhead{(4)} &\colhead{(5)} &\colhead{(6)}&\colhead{(7)}&\colhead{(8)}&\colhead{(9)}} 
\startdata  

UCM0014+1829 & 1.52$\pm$0.07 & 1.09 & 0.98 & 1.56 & 133 &  7.5 & 2.9 & 0.05 & 0.62 \\
UCM0040+0220 & 0.26$\pm$0.03 & 0.70 & 0.63 & 0.41 & 146 &  2.9 & 1.4 & 0.02 & 0.20 \\
UCM0056+0043 & 2.90$\pm$0.11 & 0.51 & 0.46 & 6.34 &  77 &  4.1 & 0.5 & 0.54 & 0.20 \\
UCM0135+2242 & 4.32$\pm$0.21 & 1.87 & 1.68 & 2.54 & 177 &  6.3 & 4.4 & 0.10 & 0.27 \\
UCM0148+2124 & 0.38$\pm$0.02 & 0.72 & 0.65 & 0.59 & 155 &  4.5 & 2.4 & 0.02 & 0.71 \\
UCM0159+2354 & 0.60$\pm$0.01 & 0.53 & 0.47 & 1.27 & 110 &  4.5 & 1.2 & 0.05 & 0.78 \\
UCM2251+2352 & 1.78$\pm$0.09 & 1.89 & 1.70 & 1.05 & 270 &  6.2 & 10.0 & 0.02 & 0.14 \\
UCM2304+1640 & 1.28$\pm$0.03 & 0.39 & 0.35 & 3.65 & 110 &  2.9 & 0.8 & 0.16 & 0.53 \\
UCM2325+2318 & 4.09$\pm$0.03 & 26.3 & 23.5 & 0.18 & 116 &  8.3 & 2.5 & 0.16 & 0.07\\
UCM2329+2512 & 1.66$\pm$0.05 & 0.25 & 0.23 & 7.29 & 144 &  2.5 & 1.2 & 0.14 & 0.60\\
UCM2351+2321 & $<0.13$ & 1.33 & 4.53 & $<0.03$ & 144\tablenotemark{a}  & 3.2  & 1.5  & $\le$0.01 & 0.31 \\
\enddata  
\tablenotetext{1}{\HI\ mass, $M(\HI)=2.356\times10^{5}~D^2~\int~S~dv$ [M$_{\odot}$].}
\tablenotetext{2}{H$\alpha$ luminosity, from Gallego {\it{et al.}} (1996), rescaled to $H_0=70$.}
\tablenotetext{3}{Star formation rate, SFR $=\frac{L_{H\alpha}}{1.12\times 10^{41}}$, from Kennicutt 1983.}
\tablenotetext{4}{Gas depletion timescale, $\tau_{gas}$=M$_{\HI}$/SFR, from Kennicutt 1983.}
\tablenotetext{5}{\HI\ rotation velocity corrected for inclination, V$_{rot}$=0.5$\times$W$_{20}/sin(i)$.}
\tablenotetext{6}{\HI\ radius, R$_{\HI}\sim$4.5$\times$R$_{eff}$, see text for derivation.}
\tablenotetext{7}{Dynamical mass corrected for inclination, M$_{dyn}$=V$^2_{rot}$R$_{\HI}/G$. }
\tablenotetext{8}{f$_{gas}$=M$_{\HI}$/M$_{Dyn}$.  Mass of molecular gas and ionized gas not included. }
\tablenotetext{9}{Mass-to-Light ratio at R$_{eff}$.  Mass is calculated from 
the velocity dispersion, $\sigma$=W$_{20}$(\HI)/3.62, without correcting for
inclination.  Light is the blue luminosity contained within R$_{eff}$ (i.e. 
half the total luminosity).}
\tablenotetext{a}{from the \HII\ width in Table 2}

\end{deluxetable}  

\newpage

\begin{figure}
\plotone{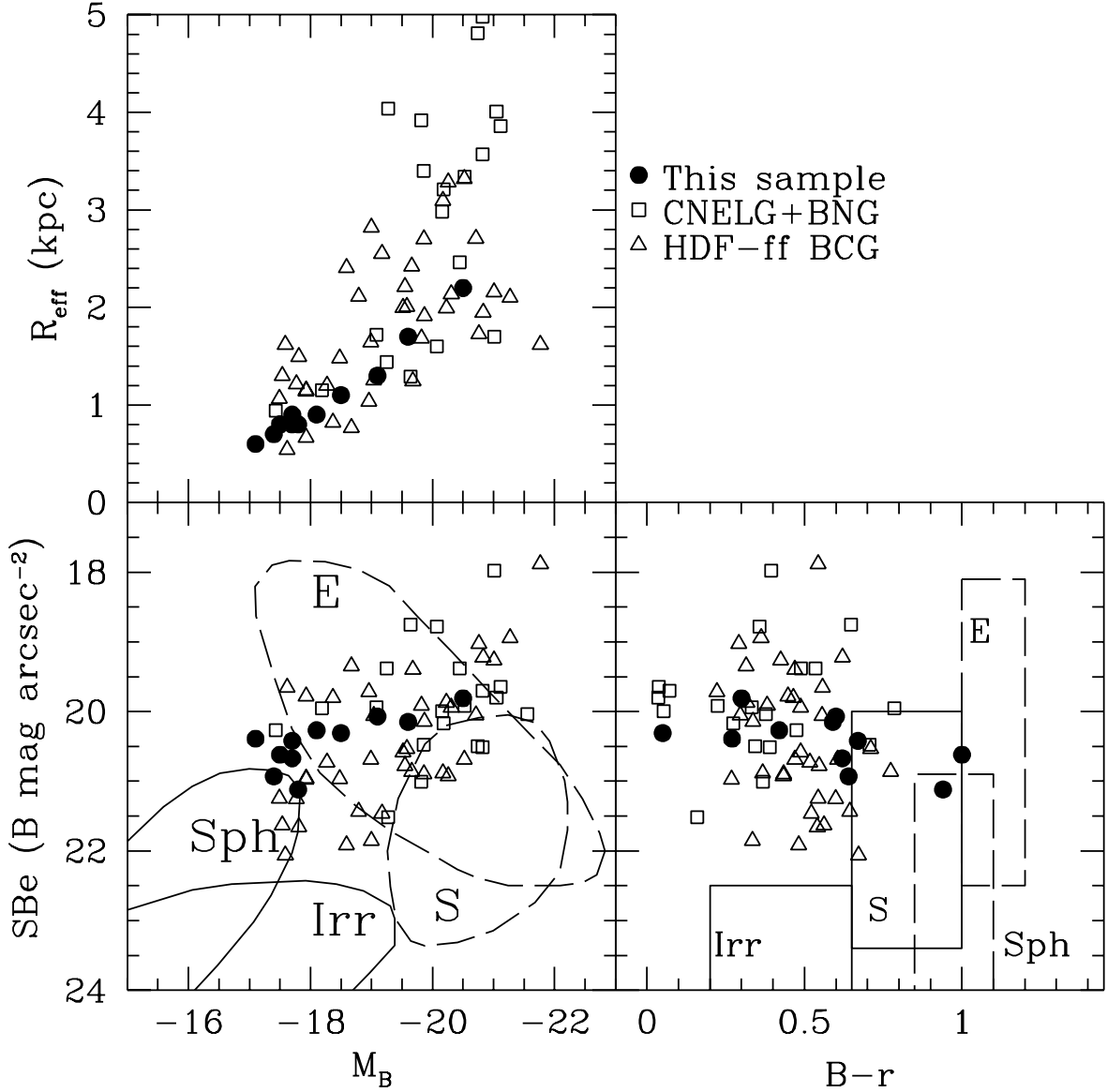}
\caption{A comparison of the selection criteria for 11 UCM sample
galaxies  (filled circles) with CNELGs and BNGs (squares) from Jangren
\etal\ (2000) and compact galaxies in the  HDF flanking fields
(triangles) from Phillips {\it{et al.}} (1997) and Guzm\'an {\it{et al.}}
(1997).  The lower left panel shows absolute magnitude vs. B-band
surface brightness and includes polygons outlining the general regions
where nearby elliptical (E), spiral (S), irregular (Irr), and
spheroidal (Sph) galaxies would lie, based on Guzm\'an \etal\ (1997).
The lower right panel shows color $B-r$ vs. B-band surface brightness
($SB_e$), with similar polygons.  Upper left shows the absolute
magnitude versus effective radius, $R_{eff}$.  The sample galaxies
have similar luminosities, surface brightnesses and $B-r$ colors as
the more distant BCGs.}
\end{figure}

\begin{figure}
\plotone{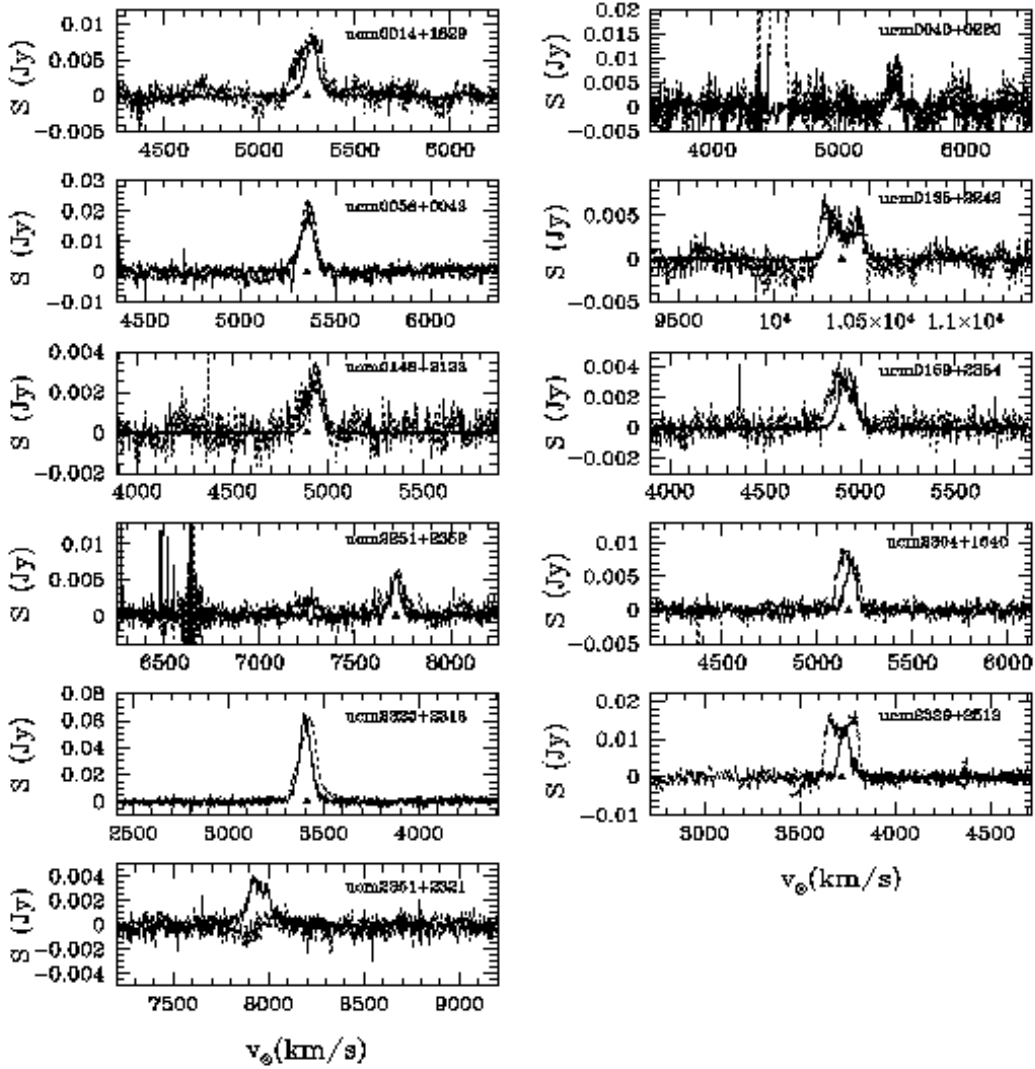}
\caption{\HI\ 21 cm spectra of the sample galaxies plotted as
Janskys versus heliocentric recession velocity defined in the optical
sense.  All but one galaxy is detected.  The dark lines are the
optical \HII\ emission-line spectra normalized to match
the \HI\ Arecibo data in grey.  The triangles mark the central
velocity of the \HI\ profile.  This figure shows 
good agreement between the systemic velocities and profile shapes in 
\HII\ and \HI, but note that the \HII\ lines are systematically more 
narrow than the \HI.}
\end{figure}

\begin{figure}
\epsscale{0.80}
\plotone{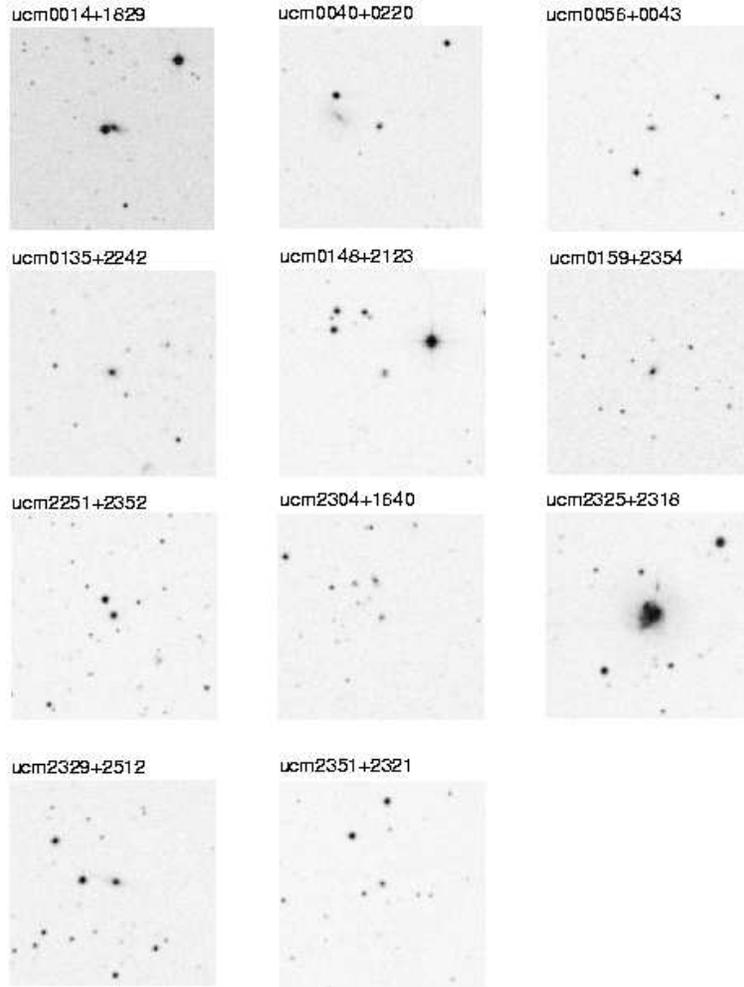}
\caption{Optical images from the Digital Sky Survey centered on the 
coordinates observed for each galaxy showing a patch of sky 3.$^\prime$8 on a 
side to match the largest dimension of the Arecibo beam at 21 cm.}
\end{figure}

\begin{figure}
\epsscale{0.70}
\plotone{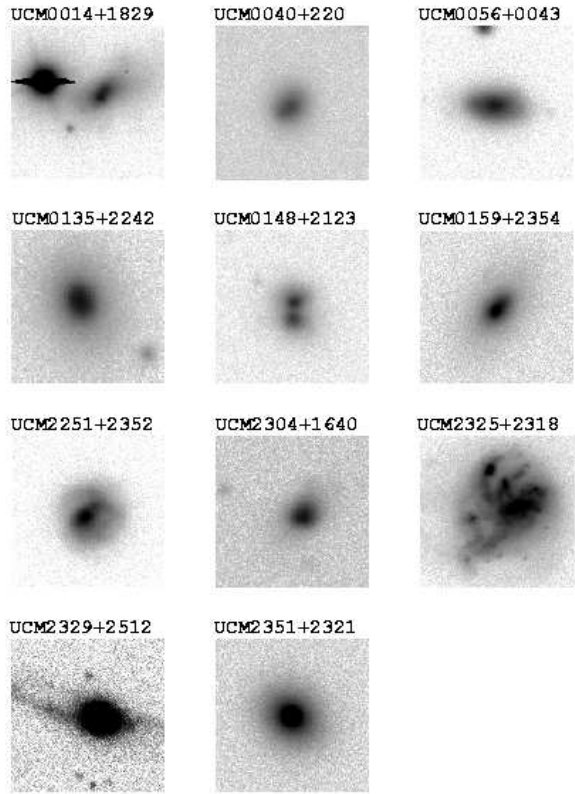}
\caption{R-band logarithmic greyscale
images from the WIYN 3.5 m telescope obtained in 1.2\arcsec\ mean seeing.
North is up, and images are 10 kpc on a side at the adopted distance of each 
source.}
\end{figure}

\begin{figure}
\plotone{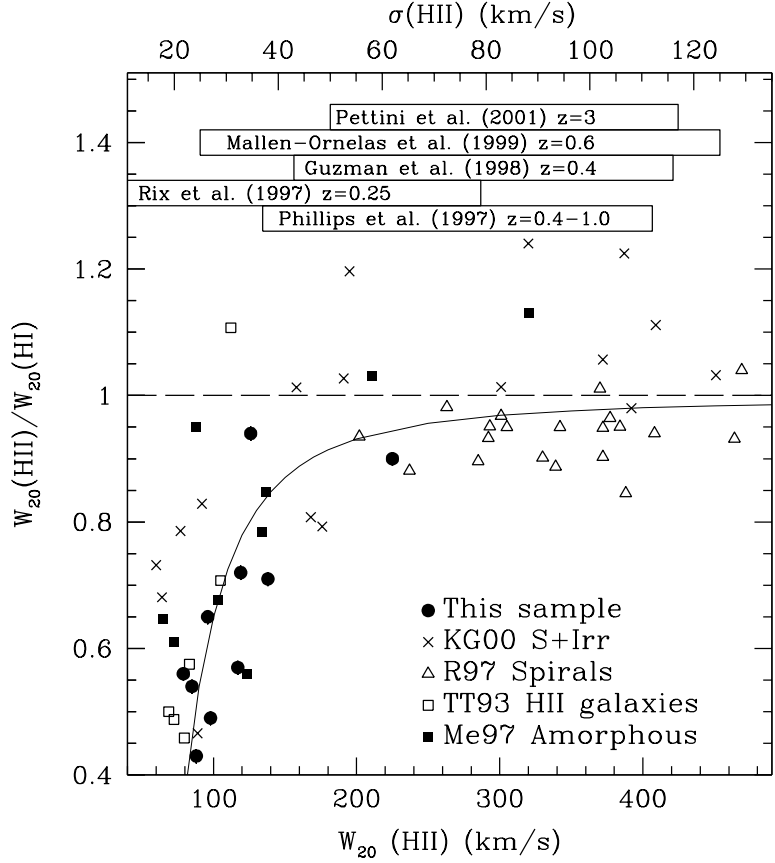}
\caption{Comparison of the observed \HII\ linewidths (i.e., not
corrected for inclination) versus the ratio of \HII-to-\HI\
linewidths.  Filled circles denote our sample, crosses denote the
sample of spiral \& irregular galaxies studied by Kobulnicky \&
Gebhardt (2000), open triangles denote the spiral sample of
Raychaudhury \etal\ (1997), and open squares are the \HII\ galaxies
from Telles \& Terlevich (1993) with \HI\ widths as measured by Smoker
\etal\ (2000).  The measured \HII-to-\HI\ ratio departs systematically
from unity for galaxies with small \HII\ linewidths.  The solid line
is an approximate fit to the data: $W_{\HII}/W_{\HI}= 1 -
5W_{\HII}^{-1} - 500W_{\HII}^{-2} - 2.5\times10^{5}W_{\HII}^{-3}$ The
boxes at the top of the figure illustrate the range of velocity widths
measured for compact galaxies in the Hubble Deep Field flanking fields
(Phillips \etal\ 1997), $z\sim0.6$ field galaxies (Mall\'en-Ornelas
\etal\ 1999), $z\sim0.25$ field galaxies (Rix \etal\ 1997), and
compact blue galaxies at $z\sim0.4$ (Guzm\'an \etal\ (1998).  Our
sample of compact blue galaxies have $W_{\HII}/W_{\HI}\sim0.6$ similar
to the ratio for \HII\ galaxies studied by Telles \& Terlevich
(1993).  This effect is capable of reducing the T-F offsets found at 
intermediate redshifts by some studies.}
\end{figure}

\begin{figure}
\plotone{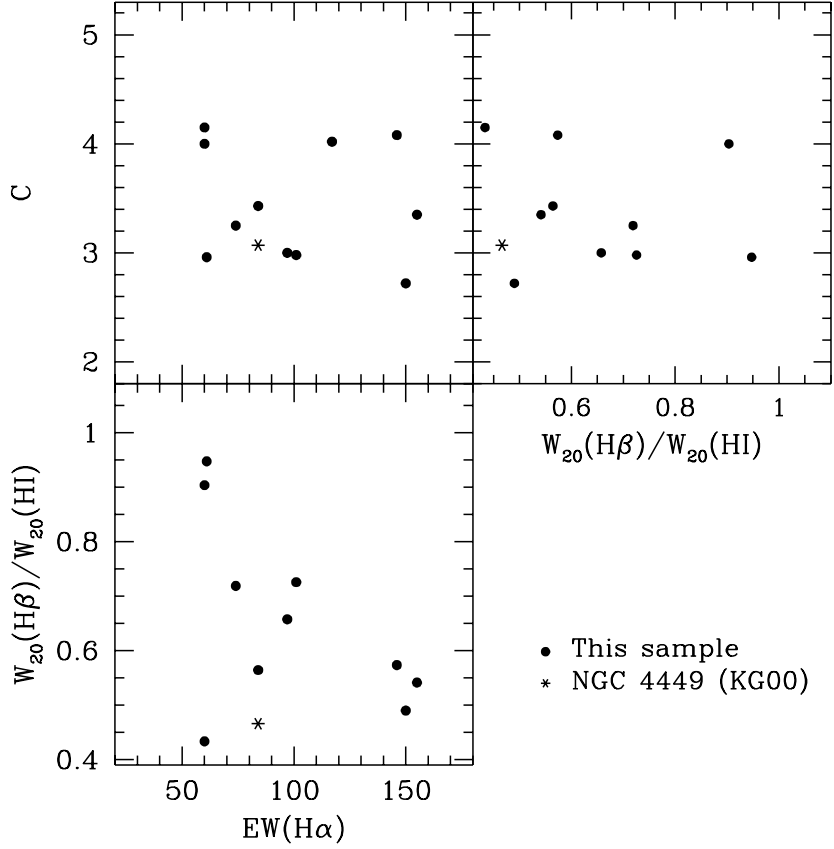}
\caption{Diagnostic parameters of the nearby galaxy sample (filled circles), 
illustrating the relation between linewidth ratio,
($W_{20}(\HII)/W_{20}(\HI)$), H$\alpha$ equivalent width, $EW(H\alpha)$, 
and concentration index, $C$.
The nearby irregular galaxy NGC 4449, marked by a star, is included as a 
comparison object.  Objects with high $EW(H\alpha)$ tend to have the 
smallest $\HII/HI$ linewidths, although the dispersion is large.  }
\end{figure}

\begin{figure}
\plotone{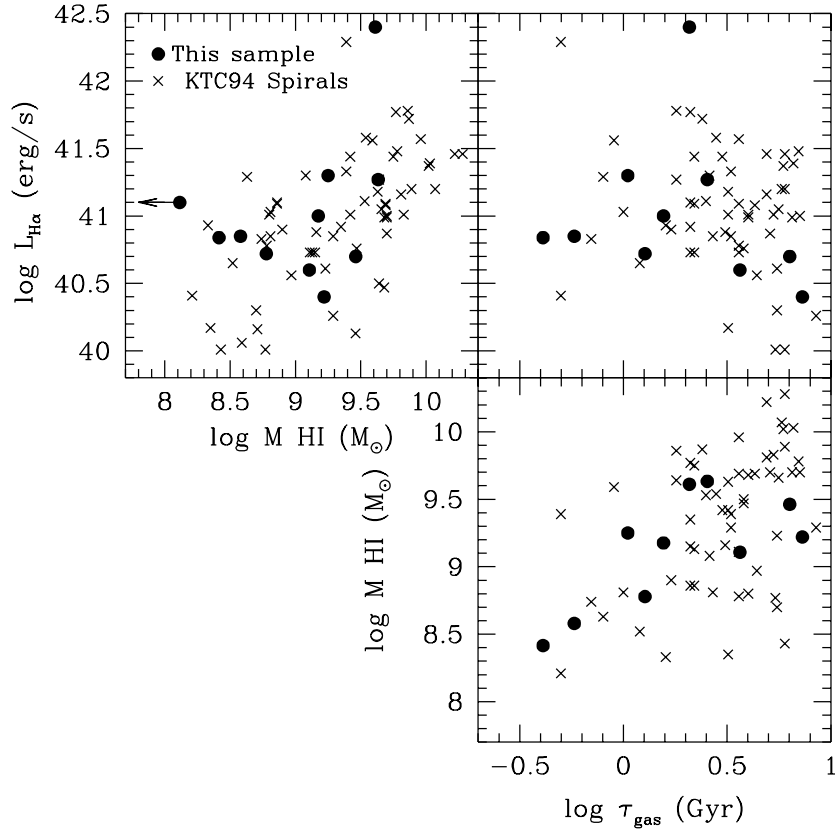}
\caption{A comparison of M$_{\HI}$ and L$_{H\alpha}$ (upper left panel), 
$\tau_{gas}$ and L$_{H\alpha}$ (upper right panel), and M$_{\HI}$ and 
$\tau_{gas}$ (lower right panel) for the UCM BCG analogs 
(filled circles) and a sample of spiral galaxies from Kennicutt 
{\it{et al.}} (1994; crosses).  It is evident that the two samples of 
galaxies occupy a similar region of parameter space in all three plots, 
although a couple of UCM galaxies do have much shorter $\tau_{gas}$ than the 
Kennicutt {\it{et al.}}  (1994) sample.  The upper limit of M$_{\HI}$ and
$\tau_{gas}$ for UCM 2351+2321 is off the left side of the plot.}
\end{figure}

\begin{figure}
\plotone{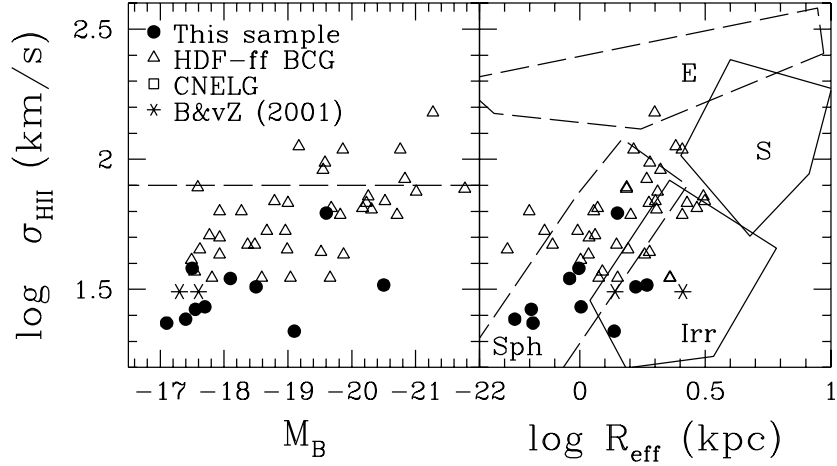}
\caption{A comparison of the sample galaxies (filled circles) with the same
intermediate-redshift BCGs (squares and triangles) from Figure 1.  
The left panel compares the velocity dispersion with absolute blue
magnitude (M$_B$). The right panel compares the emission line velocity
dispersion, $\sigma_{\HII}$, (not corrected for inclination) with the logarithm
of the effective radius (R$_{eff}$).  The velocity dispersion is measured
via the H$\beta$ line for our sample, and via a combination of H$\beta$, 
[O~II], and [O~III] linewidths for the intermediate redshift BCGs.  
Our sample galaxies have smaller velocity dispersions than the
intermediate redshift BCGs, on average.}
\end{figure}

\begin{figure}
\plotone{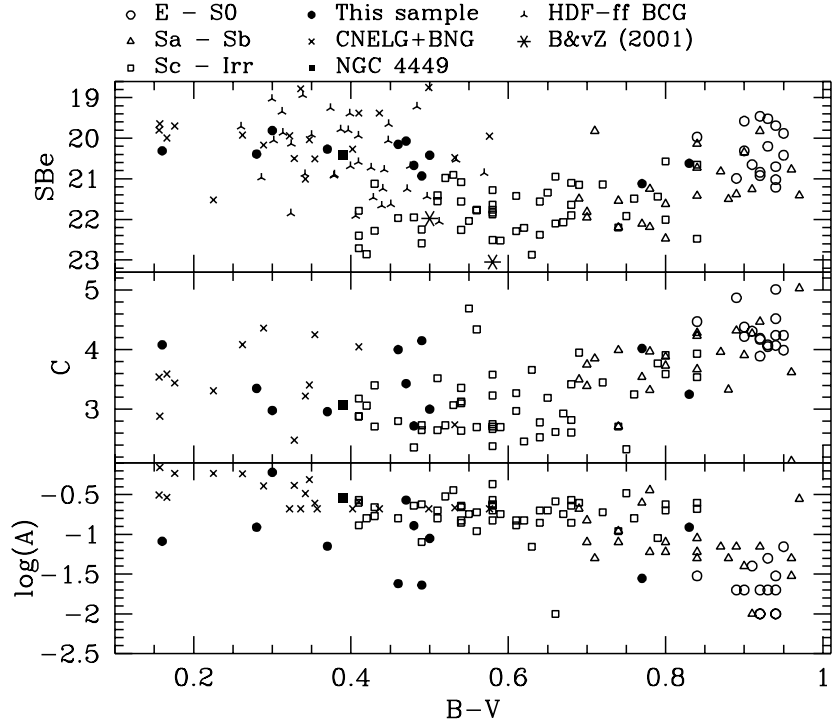}
\caption{Diagnostic parameters including
color (B-V),  B-band surface brightness
(SBe in mag $arcsec^{-2}$), concentration index ($C$), and asymmetry ($A$).  
Filled circles denote our sample, open symbols denote nearby galaxies 
from the Frei (1999) catalog taken from Bershady \etal\ (2000).  Crosses and
upside-down `Y's denote the BCG collection of Jangren \etal\ (2000).
Note that our sample of nearby compact galaxies shares a unique space
in these diagrams with the intermediate-redshift BCGs, indicating that analogs of 
the intermediate-redshift compact galaxies can be found among our sample.  }
\end{figure}

\begin{figure}
\plotone{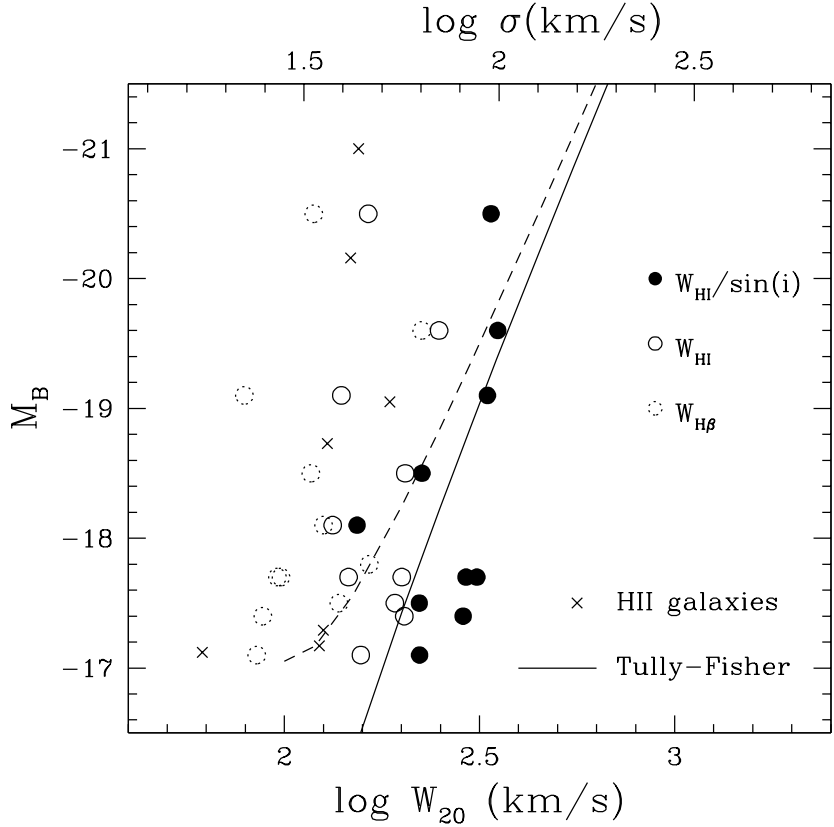}
\caption{Comparison between our sample galaxy kinematics (circles), 
\HII\ galaxies (crosses) from Telles \& Terlevich (1993) and the canonical 
Tully-Fisher relation (solid line:  $M_B=-6.86 \log(W_R)-2.27$ from 
Pierce \& Tully 1988).  The raw \HII\ velocity widths (dashed circles) 
for our sample are most consistent with the locus of the \HII\ galaxies.  
The \HI\ velocity widths (open circles) are closer to the Tully-Fisher 
relation.  But the inclination-corrected \HI\ velocity widths (filled circles) 
are consistent with the Tully-Fisher without any luminosity offsets. The 
dashed curve shows how the T-F relation would be modified if measured using 
inclination-corrected optical line-widths, based on the relation adopted in 
Fig. 5.}
\end{figure}

\end{document}